\documentclass[pdflatex,sn-nature]{sn-jnl}

\usepackage{graphicx}%
\usepackage{multirow}%
\usepackage{amsmath,amssymb,amsfonts}%
\usepackage{amsthm}%
\usepackage{mathrsfs}%
\usepackage[title]{appendix}%
\usepackage{xcolor}%
\usepackage{textcomp}%
\usepackage{manyfoot}%
\usepackage{booktabs}%
\usepackage{algorithm}%
\usepackage{algorithmicx}%
\usepackage{algpseudocode}%
\usepackage{listings}%

\usepackage[version=4]{mhchem}%
\usepackage{siunitx}%
\usepackage{physics}%
\usepackage{braket}%
\usepackage{float}%

\theoremstyle{thmstyleone}%

\theoremstyle{thmstyletwo}%

\theoremstyle{thmstylethree}%

\makeatletter
\renewcommand{\ps@plain}{%
  \let\@mkboth\@gobbletwo%
  \let\@oddhead\@empty%
  \let\@evenhead\@empty%
  \def\@oddfoot{\reset@font\rmfamily\hfil\thepage\hfil}%
  \def\@evenfoot{\reset@font\rmfamily\hfil\thepage\hfil}%
}
\makeatother
\begin{document}

\pagestyle{plain}

\title[Quantum Cell Monitoring for Live-Cell Imaging]{Multi-day single-cell magnetic imaging of living cancer cells using Quantum Cell Monitoring}

\author[1]{\fnm{Andre} \sur{Pointner}}
\author[2]{\fnm{Sarah} \sur{Belasi}}
\author[2]{\fnm{Regine} \sur{Schneider-Stock}}
\author*[1]{\fnm{Roland} \sur{Nagy}}\email{roland.nagy@fau.de}

\affil[1]{\orgdiv{Institute of Applied Quantum Technologies}, \orgname{Friedrich-Alexander-Universit\"{a}t Erlangen-N\"{u}rnberg}, \orgaddress{\city{Erlangen}, \country{Germany}}}
\affil[2]{\orgdiv{Experimental Tumor Pathology}, \orgname{University Hospital Erlangen}, \orgaddress{\city{Erlangen}, \country{Germany}}}

\abstract{Live-cell imaging of cellular and tissue samples relies predominantly
on optical modalities, such as fluorescence, confocal, and multiphoton
microscopy, in which cellular information is encoded in emitted photons. Photons
scatter and attenuate in biological samples, restricting depth resolution, while
the excitation light required to produce them induces photodamage that perturbs
the very biology under observation. Magnetic fields, by contrast, traverse
biological samples without attenuation at cellular length scales, but the
imaging modality that extracts the encoded information, magnetic resonance
imaging, cannot resolve individual cells. Here we present Quantum Cell
Monitoring (QCM), a widefield nitrogen-vacancy magnetometry platform that
encodes single-cell information in stray magnetic fields rather than emitted
photons and reads it out at subcellular resolution. Integrated with a
specialized live-cell incubator, QCM tracks magnetically labeled living cancer
for up to \SI{72}{\hour} through their individual magnetic field signatures. It
captures single-cell mitosis and morphological remodeling, as well as cell
cluster formation, splitting, and migration even within dense regions where
brightfield contrast collapses. By shifting the carrier of single-cell
information from propagating optical photons to static magnetic stray fields,
QCM decouples cell from cellular optical accessibility and auto-fluorescence,
enabling sustained live-cell observation in regimes inaccessible to optical
microscopy.}


\maketitle

\section*{Introduction}
Fluorescence microscopy, the prevailing method for live-cell imaging, faces
three coupled limitations that together impede multi-day single-cell
observation: (i) durability of the label, (ii) strong excitation light directly
incident on the cell, and (iii) scattering of the fluorescence emission
restricting depth resolution. While photobleaching of small-molecule dyes under
extended imaging durations can be avoided by genetically encoded labels such as
photostable fluorescent proteins \cite{Hirano2022,Laissue2017}, the excitation
light needed to drive these labels affects cell behavior well below the
threshold of cell damage. As a result, mitotic delay and altered migration
speeds are observed at doses routinely incurred during multi-day imaging
\cite{DelRosario2025,Icha2017}. Taken together, these limitations preclude
diffraction-limited fluorescence imaging at biologically relevant depths, even
with multiphoton excitation \cite{Yoon2020}. These constraints are not
independent symptoms but consequences of using photons as the carrier of
cellular information. Mitigating one typically increases the others
\cite{Mela2024}, imposing fundamental limitations on the applicability of this
technique for extended single-cell observation. Magnetic resonance imaging
(MRI), by contrast, replaces photons with magnetic fields as the information
carrier, exploiting their ability to penetrate biological tissue without
attenuation or interaction with the sample at cellular length scales, but lacks
the spatial resolution required to resolve individual cells
\cite{Shapiro2004,Cheng2023,Heyn2006}. As a result, events at the single-cell
length scale, such as division, morphological remodeling, and short-range
migration inside live tissue remain inaccessible to existing imaging techniques
over multi-day timescales. Closing this gap requires both a magnetic label that
encodes single-cell information into magnetic stray fields and a sensor that
detects those fields at the cellular length scale under live-cell conditions.

Such a label is provided by the selective labeling of cells with
superparamagnetic iron oxide nanoparticles (SPIONs) \cite{Friedrich2022}, which
provide a stable and biocompatible \cite{Feng2018, Hoskins2012}
magnetic contrast whose faint stray fields can in principle be detected to
identify and track individual cells. These labels can be
selectively functionalized for specific cellular targets, based on the desired
application through antibody targeting. To date, however, no imaging modality
has been demonstrated that resolves these stray fields at the single-cell level
under long-term live-cell conditions. Such measurements require a detection
method that simultaneously delivers high magnetic field sensitivity, subcellular
spatial resolution and full compatibility with live-cell culture conditions.
Nitrogen-vacancy (NV) centers in diamond meet these requirements, combining high
magnetic field sensitivity with diffraction-limited spatial resolution and an
inert, biocompatible sensing substrate \cite{Hollendonner2023}. This
combination, in principle, unites the magnetic-field information carrier with
the single-cell resolution of optical microscopy, closing the
carrier-versus-resolution tradeoff that has so far separated the two modalities.

Widefield NV magnetometry has been demonstrated to image magnetically labeled
single cells \cite{LeSage2013}, to resolve internalized SPIONs in fixed samples
\cite{Mathes2024}, and to map excised tumor tissue \cite{Chen2022}. While these
prior pioneering approaches successfully resolved static biomarker expression in
short proof-of-principle demonstrations with single time point \cite{Glenn2015}
or short-term imaging \cite{Davis2018} of living cell samples, sustained
longitudinal imaging of living cells has remained fundamentally elusive. The
critical bottleneck is that multi-day live-cell diamond magnetometry requires
the complex orchestration of non-invasive continuous microwave delivery, high
sensitivity spin readout under strictly minimized photodamage, and a specialized
incubation environment. Consequently, monitoring single-cell dynamics like
proliferation or mitosis through purely magnetic signatures over biologically
relevant timescales has never been achieved.

Here, we present Quantum Cell Monitoring (QCM), a self-contained
quantum-biological platform built around a custom-engineered live-cell
incubation architecture \cite{Pointner2025incubator}. Unlike conventional
commercial chambers, this proprietary hardware is co-optimized for fluidic and
optical performance to solve a dual challenge: it suppresses parasitic stray
light to ensure efficient light coupling at the sensing interface, while
simultaneously maintaining long-term cell viability by providing a stable
nutrient supply and strict homeostatic regulation. This innovation breaks the
long-standing barrier of thermal and phototoxic cell degradation during
continuous diamond magnetometry, transforming widefield NV sensors from a static
endpoint tool into an autonomous, longitudinal live-cell imaging modality.

We resolve single-cell migration, proliferation and magnetic-signal dynamics in
a monolayer culture over up to \SI{72}{\hour}. Together, these results establish
QCM as a live-cell imaging technique in which single-cell identity is carried by
stray magnetic fields rather than emitted photons, decoupling cell monitoring
from the optical accessibility of the sample and resolving single-cell dynamics
on living samples over multi-day timescales.

\section*{Results}
\subsection*{Widefield nitrogen-vacancy center magnetometry for live-cell imaging}\label{sec:setup}

\begin{figure}[H]
    \centering
    \includegraphics{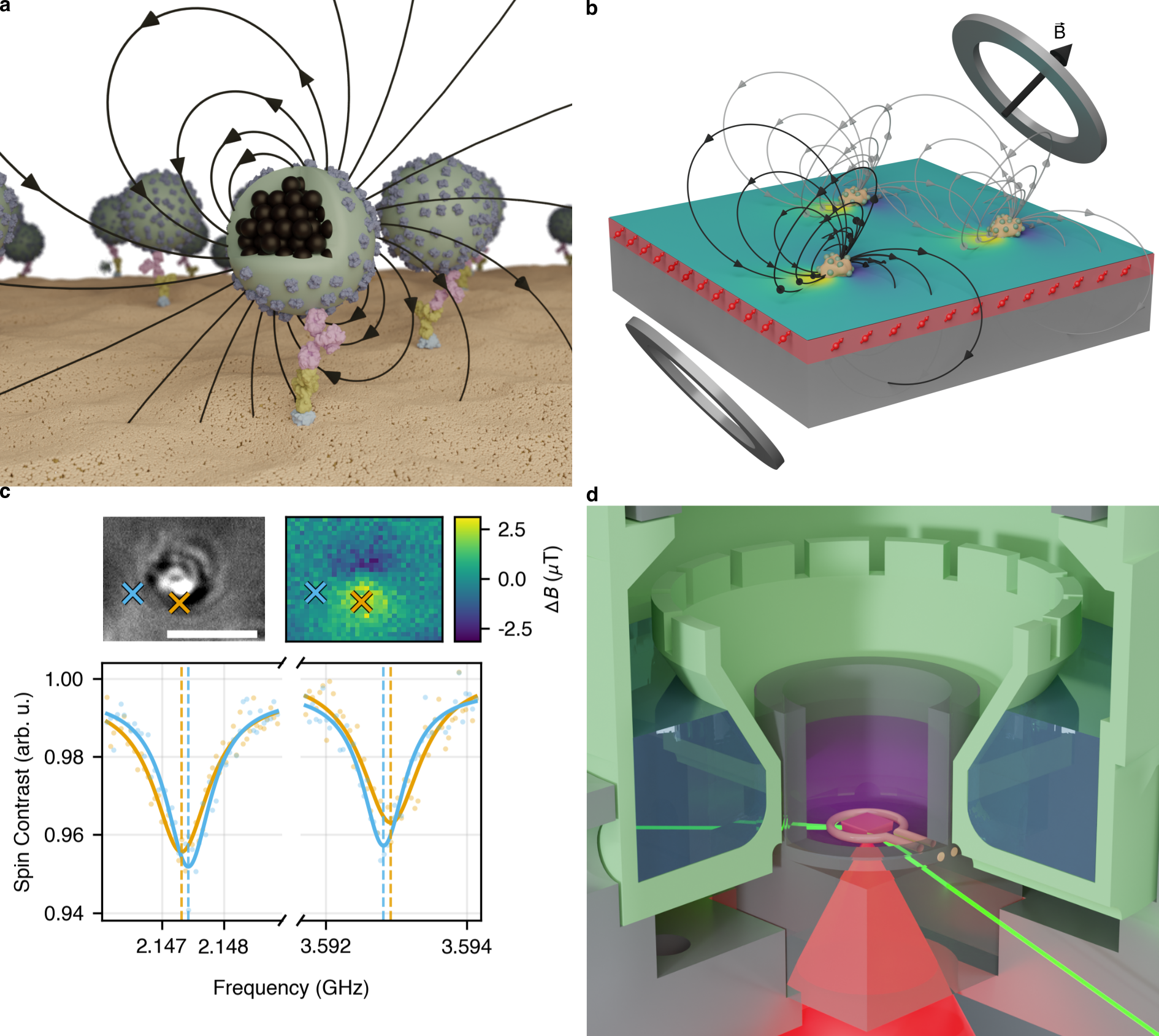}
    \caption{\textbf{Widefield NV center magnetic field microscope for long-term live-cell imaging.}
    \textbf{a}, Illustration of the magnetic labeling protocol. SPIONs (black)
    are coated in a polymer shell (green) and functionalized with Streptavidin.
    The primary antibody (yellow) attaches to the EpCAM surface marker (blue) of
    the cell membrane (brown). The biotinylated secondary antibody (pink)
    attaches to the primary antibody and provides a binding site for the
    streptavidin on the SPION surface, resulting in the attachment of SPIONs to
    available EpCAM surface markers. Magnetization of the sample by a pair of
    permanent magnets causes the superparamagnetic particles to generate a
    magnetic dipole field (black field lines). \textbf{b}, Render of the sample
    assembly. The magnetically labeled cells (brown with SPIONs illustrated in
    green) are seeded on the diamond's NV layer surface (red). A pair of
    permanent ring magnets (gray) create a magnetizing bias field $\mathbf{B}$
    aligned with a $\lbrack 111\rbrack$ axis of the diamond. The magnetic dipole
    signature of the cells is detected by the NV centers through widefield ODMR
    measurements (color map). \textbf{c}, NV Center spin resonance shifts
    induced by local magnetic dipole fields from a magnetically labeled cell
    (brightfield image left in black and white contrast). The surface-bound
    SPIONs cumulatively produce a magnetic field signature that shifts the NV
    center's resonance frequencies from the magnetic bias field resonance. This
    shift (blue: on the cell, orange: off the cell) allows reconstruction of the
    local absolute magnetic field value affecting the NV centers of the
    respective pixels. Background subtraction of the magnetic bias gradient by a
    rolling ball algorithm reveals the magnetic dipole signatures of the
    individual cells. The selected pixels observe a shift of the resonance lines
    by $\approx$\SI{134}{kHz}, resulting in an absolute magnetic field
    difference between the two pixels of \SI{4.79}{\mu T}. After background
    subtraction, the magnetic field signature of the proximal cell (magnetic
    field image on the right) reveals a \SI{3.49}{\mu T} magnetic field strength
    between the two pixels. \textbf{d}, Render of the specialized incubation
    system integrated with the widefield NV microscope. Green, incubation
    chamber enclosing the cell culture atmosphere; red, the $2\times 2$ \si{mm}
    diamond sample emitting into the objective below; green line, excitation
    laser; orange, microwave antenna in proximity to the diamond sample, which
    is mounted on a $12\times 12 \times 8$ \si{mm} glass cube for total internal
    reflection (TIR) excitation of the NV layer.}
    \label{fig:figure_1}
\end{figure}

We use commercially available \SI{20}{\nano\meter} core size SPIONs with a
polymer shell that is functionalized with streptavidin, allowing for
antibody-based targeting of the particles to cell surfaces
(Fig.~\ref{fig:figure_1}a). We target the epithelial cell adhesion molecule
(EpCAM), a surface marker commonly expressed in epithelial cancers, including
the HT29 colorectal cancer cell line used in this study. Optimizing the labeling
protocol, we achieve saturated binding of SPIONs to the cell surface, resulting
in a uniform distribution of magnetic labels across the cell membrane. Previous
work \cite{Pointner2025} has shown, that fully saturated HT29 cells can bind up
to \SI{2.5}{\pico\gram} of SPIONs, which corresponds to an approximate (using
the mean density of maghemite/magnetite and assuming single-particle linkage)
number of $N\approx 11.85\times10^{4}$ SPIONs on the cell surface. The
SPION-labeled cells are seeded directly onto the diamond sensor, hosting a
\SI{1}{\mu\meter} thick ensemble layer with a NV center density of \SI{1}{ppm}
(Fig.~\ref{fig:figure_1}b). The cells adhere to the diamond surface, mediated by
a Poly-L-Lysine (PLL) coating that improves cell adhesion. An external magnetic
bias field magnetizes the SPIONs on the cell surface
(Fig.~\ref{fig:figure_1}a,b) and subsequently induces a cumulative magnetic
dipole field for each individual cell approximately following a sum of magnetic
dipoles:
\begin{equation}
\mathbf{B}_{\mathrm{c}}(\mathbf{r}) = \frac{\mu_0}{4\pi} \sum_{i=1}^{N} \left( \frac{3(\mathbf{m} \cdot \mathbf{d}_i)\mathbf{d}_i}{|\mathbf{d}_i|^5} - \frac{\mathbf{m}}{|\mathbf{d}_i|^3} \right)\text{,}
\label{eq:B}
\end{equation}
where $\mathbf{d}_i = \mathbf{r} - \mathbf{r}_i$ is the vector from the $i$-th
 SPION at $\mathbf{r}_i$ to position $\mathbf{r}$,
 $\mathbf{B}_{\mathrm{c}}(\mathbf{r})$ is the magnetic field vector at position
 $\mathbf{r}$, $N$ is the number of SPIONs on the cell surface, $\mu_0$ is the
 vacuum permeability, $\mathbf{m}$ is the magnetic moment of the $i$-th SPION.
 The resulting magnetic field encodes information about the position and shape
 of the cell in the form of shifting $\mathbf{d}_\mathrm{i}$ and subsequently
 the magnetic field signature of the cell relative to the NV center layer
 \cite{Pointner2025}. This magnetic dipole field of each cell causes local
 deviations of the much larger magnetic bias field, which in turn locally shifts
 the NV center's spin resonance frequencies based on the electron Zeeman effect.
 The resulting local shift from the magnetic bias field at location $\mathbf{r}$
 is $\Delta f(\mathbf{r}) = \gamma B_{c,\parallel}(\mathbf{r})$, where $\gamma =
 \SI{28}{kHz/\mu T}$ and $B_{c,\parallel}(\mathbf{r})$ is the local magnetic
 field generated by the cell (Eq.~\ref{eq:B}) projected on the interrogated NV
 center axis $\lbrack 111 \rbrack$. For a cell surface saturated with SPIONs,
 this results in a magnetic dipole field in the $\mu T$ range, which in turn
 causes shifts of the NV center's spin resonance frequencies on the order of a
 few \SI{100}{kHz} (Fig.~\ref{fig:figure_1}c). In order to detect these faint
 signals, laser excitation optically cycles the NV centers continuously, and a
 dual-tone microwave signal with a $^{15}$N hyperfine splitting governed
 separation of \SI{3.05}{MHz} is swept in frequency across both the NV center's
 $\ket{m_s=0,m_I=\pm1/2}\rightarrow \ket{m_s=\pm 1, m_I=\pm1/2}$ spin
 transitions. This oscillating magnetic field induces resonance-conditioned spin
 state driving. The convolution of this dual-tone microwave signal with the
 electron spin transitions results in three distinct resonance features in the
 ODMR signal. From these, the center feature representing the resonance where
 both microwave tones are resonant with their respective hyperfine transitions
 is sampled and evaluated (Fig.~\ref{fig:figure_1}c). This approach improves
 sensitivity by a factor of 1.4 due to the higher contrast achieved by
 simultaneously driving both hyperfine transitions of the $^{15}$N-coupled
 electron spin. Interleaved frames, with and without the microwave signal
 present, are captured by the camera to compute the spin contrast
 $C=1-(I_{on}/I_{off})$ governed by the spin-state-dependent emission intensity
 of the NV centers. Evaluating this contrast on a pixel-by-pixel basis with a
 spatial resolution of \SI{1.13}{\mu m ^2} per pixel for our $4\times 4$ binned
 sensor allows sampling of the resonance features of the spin transitions in
 this widefield continuous-wave optically detected magnetic resonance (ODMR)
 measurement. Thus, we capture the resulting electron Zeeman shift caused by
 both the magnetizing bias field and the magnetic dipole fields locally resolved
 across the field-of-view (FoV) (Fig.~\ref{fig:figure_1}b,c). Our hyperfine
 resolved ODMR measurement achieves a linewidth for the dual-tone driving of
 $\approx$ \SI{1}{MHz}. Extracting the electron Zeeman shifts for each pixel of
 the captured ODMR data, results in a map of absolute magnetic field values
 $B_{\parallel} = B_{BG,\parallel} + B_{c,\parallel}$, where $B_{BG,\parallel}$
 is the magnetic bias field along the NV $\lbrack 111 \rbrack$ axis. Finally, we
 subtract the background magnetic bias field gradient using the rolling ball
 method \cite{Sternberg1983}. Due to their localized size, the magnetic field
 signatures of individual cells $B_{c,\parallel}$ are preserved by the
 background subtraction method, resulting in clear signals on the imaged sensor
 surface \cite{Pointner2025,LeSage2013,Glenn2015}. Cells were allowed to adhere
 for \SI{8}{\hour} to the diamond surface before the first ODMR measurement. Due
 to the small size of the incubation volume, the culture medium was refreshed in
 \SI{8}{\hour} intervals, to maintain nutrient supply. To sustain cell viability
 over the multi-day experiment while retaining ODMR capability, we constructed a
 specialized incubation system (Supplementary Information Sec.~S1) around the
 diamond sample inside our widefield microscope (Fig.~\ref{fig:figure_1}d)
 \cite{Pointner2025incubator}. The incubator is placed between the bias magnets
 above the microscope objective and encloses the diamond sample, providing a
 closed-off atmosphere for the living samples. The system stabilizes
 temperature, humidity and CO$_2$ at the diamond surface (Supplementary
 Information Sec.~S1) and provides the optical and microwave access required for
 ODMR. By establishing these precise environmental conditions directly on the
 sensing substrate, this platform serves as an essential enabling technology
 that transcends the static, fixed cell imaging paradigm of prior diamond
 magnetometry \cite{Glenn2015} into a continuous, multi-day live-cell
 observation window. Consequently, individual cells can now be resolved and
 tracked in the magnetic field channel, allowing observation of their dynamic
 behavior over the course of the experiment. The ODMR measurement protocol was
 optimized to minimize photodamage (Supplementary Information Sec.~S2) through
 evanescent wave coupling and stray light by optimizing the laser power in a TIR
 configuration to \SI{25}{\milli\watt} over \SI{5}{\min} acquisition within our
 $\varnothing$\SI{376}{\mu\meter} FoV. At this operating point, our per-pixel
 magnetic field sensitivity of $\approx \SI{2.5}{\mu\tesla\per\sqrt\hertz}$
 remains sufficient to resolve the stray fields of individually labeled cells.
 This operating regime enables repeated ODMR acquisition over multiple days,
 exploiting the inherent photostability of magnetic labels to track individual
 cells on timescales inaccessible to fluorescence-based methods
 \cite{Friedrich2022}. The combination of the specialized incubation system
 \cite{Pointner2025incubator} and optimized ODMR protocol thus establishes a
 platform for long-term live-cell magnetometry, enabling subsequent experiments
 on monitoring cell dynamics.

\subsection*{Long-term monitoring of cancer cells in monolayer}\label{sec:monolayer}

To demonstrate sustained cell monitoring, we observed SPION-labeled HT29 cells
across multiple long-term experiments spanning up to \SI{72}{\hour}. Taking
advantage of the unique permanence of magnetic labels which, unlike fluorophores
do not photobleach, the QCM sampling interval was set to \SI{8}{\hour}. This
spacing was strategically chosen to match the slow diffusion kinetics of
adherent single cells \cite{Hohmann2022} and their \SIrange{24}{72}{\hour}
doubling time \cite{ForgueLafitte1989}, thereby minimizing unnecessary laser
induced grid interrogation while fully capturing cellular functions like
proliferation and migration. 

\begin{figure}[H]
    \centering
    \includegraphics{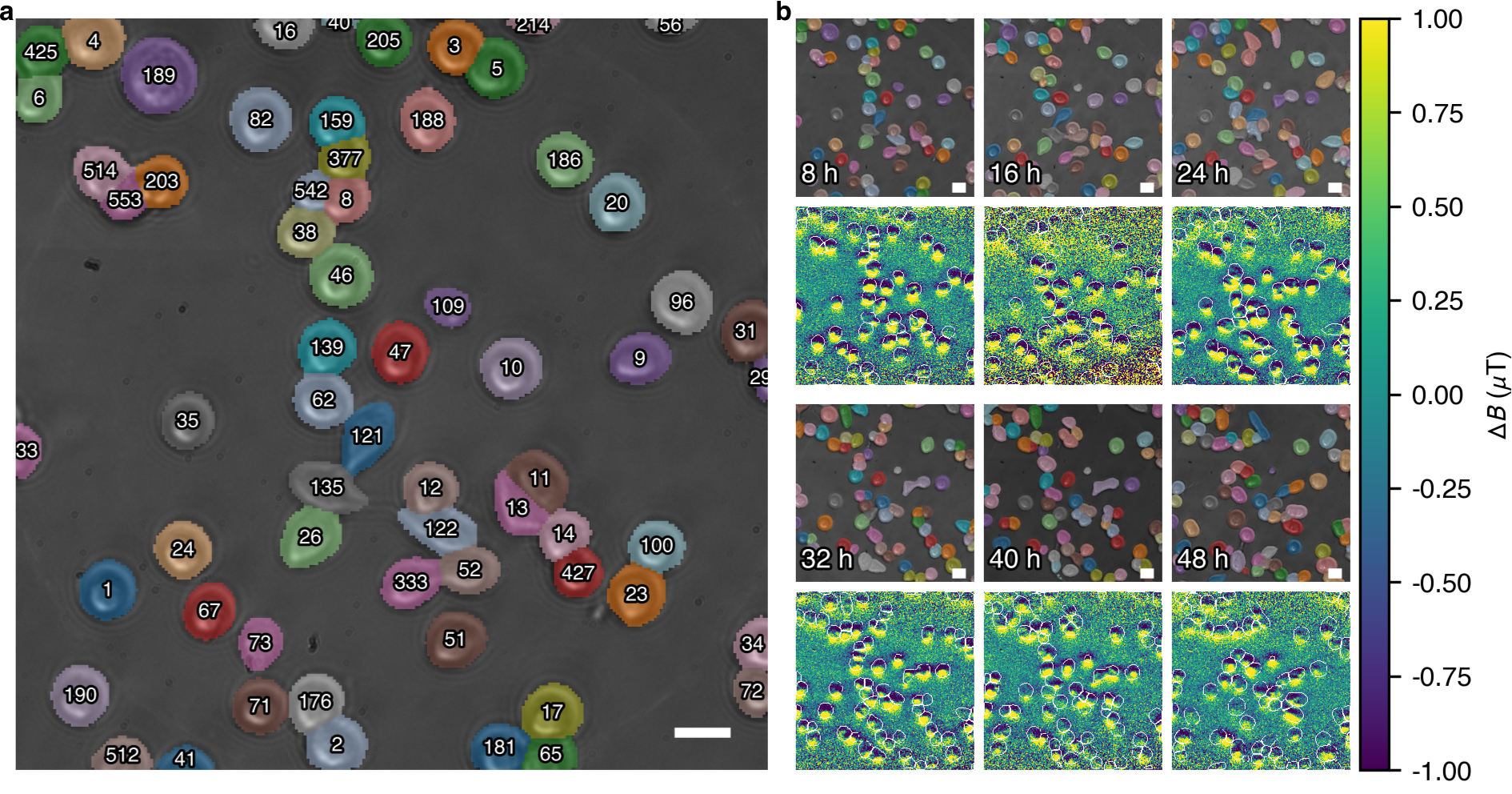}
    \caption{\textbf{Multi-day correlated brightfield and magnetic field imaging
    of SPION-labeled HT29 cells over a total of \SI{48}{\textbf{\hour}}
    timespan.} \textbf{a}, Full FoV brightfield image showing adherent cells
    with an average diameter of \SI{20}{\mu\meter} on the diamond surface after
    \SI{8}{\hour} seeding. Color overlays indicate segmented cell boundaries
    with assigned cell identities. \textbf{b}, Time series of correlated
    brightfield (top) and magnetic field (bottom) images at the ODMR measurement
    time points spanning the \SI{48}{\hour} timeframe. Colored overlays remain
    consistent. Cell identities are omitted for visual clarity. Individual
    magnetic dipole signatures of cells are visible, and cell outlines are
    overlaid based on the brightfield segmentation mask. All scale bars:
    \SI{20}{\micro\meter}.}
    \label{fig:figure_2}
\end{figure}

Concurrently, strictly to establish an independent, high-resolution ground truth
for the initial algorithmic verification of this prototype, low-intensity
brightfield reference frames were captured at 2 min intervals. As mathematically
validated later by our mean-squared displacement analysis
(Fig.~\ref{fig:figure_2}a), this high-frequency optical tracking is an
overdetermined baseline, confirming that the 8-hour magnetic channel is entirely
self-sufficient for identity retention. These optical co-registrations were
processed via Cellpose \cite{Stringer2021} and Bayesian tracking
\cite{Ulicna2021} purely to validate the analytical accuracy of our magnetic
field maps, proving that QCM maintains absolute identity retention across days
without relying on continuous optical intervention (Fig.~\ref{fig:figure_2}b).

Using the same camera for both brightfield imaging (first and third row in
Fig.~\ref{fig:figure_2}b) and corresponding ODMR measurements (second and fourth
row in Fig.~\ref{fig:figure_2}b), allows direct comparison between the two
imaging methods, by assigning cell identities across both modalities and
outlining the exact cell position within the magnetic field maps. Performing
ODMR measurements and subsequent reconstruction of the magnetic field across the
FoV provides a magnetic field map of the camera's FoV. Magnetically labeled
cells appear as characteristic dipole signatures in the magnetic field channel
(Fig.~\ref{fig:figure_2}b), arising from the projection of each stray field onto
the interrogated NV axis. This unique magnetic footprint allows locating
individual cells across the FoV. As fluorescence microscopy and subsequently
standard biological viability assays are not possible in our experimental setup,
we instead opt for a behavioral observation assessment of cell viability. Over
\SI{72}{\hour}, cells migrated, proliferated and underwent morphological
changes, confirming that the measurement protocol sustains normal cell culture
conditions. 

These results demonstrate that QCM achieves robust single-cell monitoring within
dynamic cellular environments without compromising cell viability. Notably, this
represents the first instance in quantum sensing where living biological
specimens have been continuously interrogated via diamond magnetometry over
multi-day timescales while remaining fully viable. Since the early
demonstrations of widefield magnetic imaging on fixed cells \cite{Pointner2025},
achieving long-term sample survival under sustained quantum control sequences
has remained a major barrier. By demonstrating that normal proliferation, and
migration can be tracked continuously over \SI{72}{\hour}, these findings
validate QCM not only as a non-invasive monitoring tool for oncology but as a
fundamental breakthrough in quantum-biological sensing.

\subsection*{Magnetic imaging of single-cell dynamics}

\begin{figure}[H]
    \centering
    \includegraphics{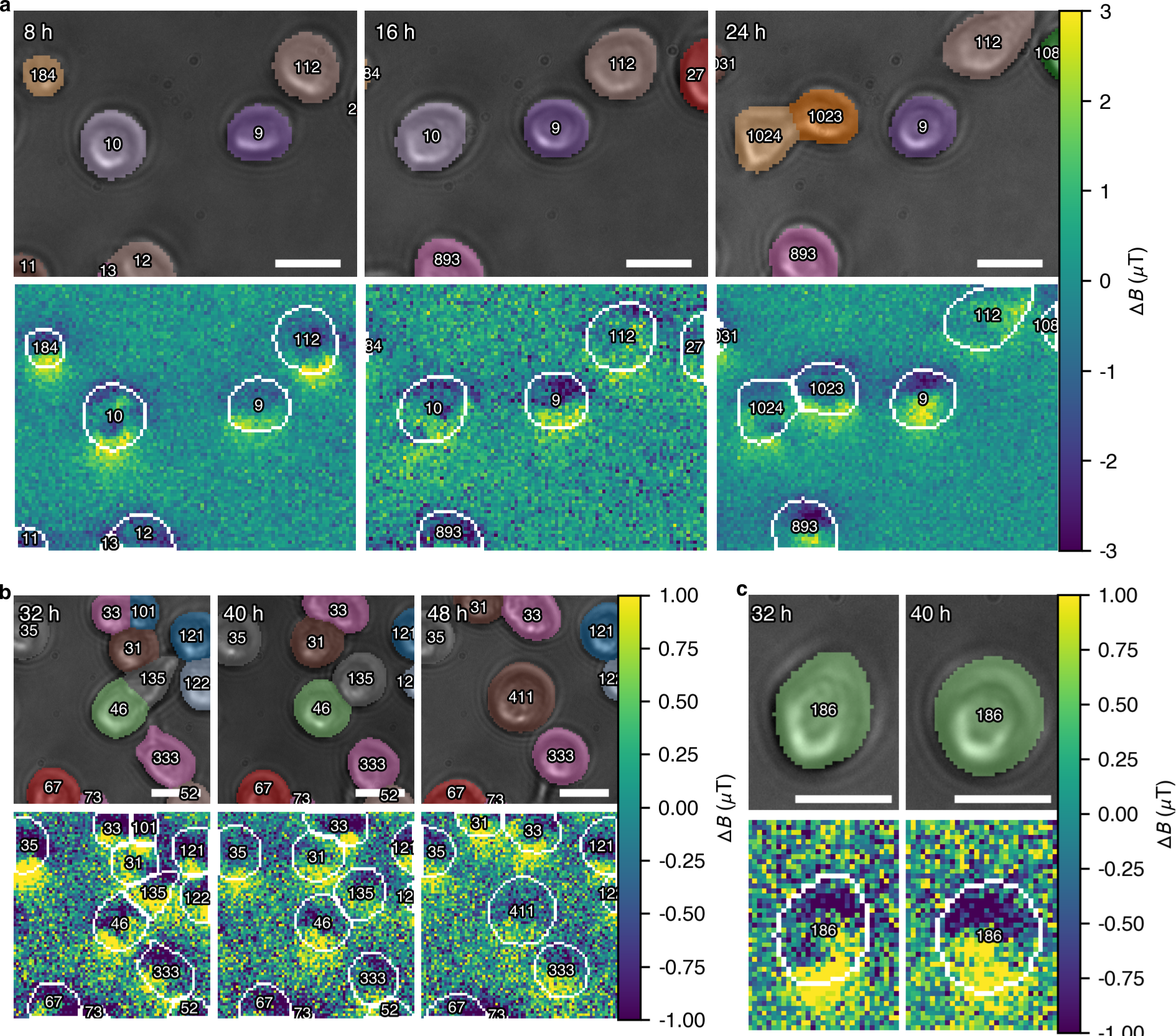}
    \caption{\textbf{Selected ROIs showing observable single cell behavior instances spanning mitosis, entosis and morphological changes.}
    \textbf{a}, Cell \#10 undergoing mitosis resulting in daughter cells \#1023
    and \#1024. The magnetic dipole is clearly split up into two distinct
    magnetic field signatures. \textbf{b}, Cell \#46 and \#135 undergo either
    tumor cell cannibalism or entosis. Observable in the merging of magnetic
    field signatures to cell \#411. \textbf{c}, Cell \#186 undergoes
    morphological changes by stretching along the vertical direction. This is
    reflected in an elongation and partial compensation of the magnetic dipole
    signature, as the positive and negative lobes of the dipole field cancel
    out over a larger area due to a change in shape of the cell surface and thus
    of the individual positions of the contributing SPIONs (Eq.~\ref{eq:B}). All
    scale bars: \SI{20}{\micro\meter}.}
    \label{fig:figure_3}
\end{figure}

The subcellular resolution of the magnetic field imaging modality allows direct
observation of single-cell dynamics over the course of a \SI{48}{\hour}
experiment. For example cell mitosis is directly observable in the splitting of
magnetic dipole signatures of individual cells into two daughter cells
(Fig.~\ref{fig:figure_3}a). Specifically, the mother cell \#10 separates into
two distinct cells clearly visible in the brightfield imaging, as well as the
magnetic field map at the \SI{24}{\hour} time point. The daughter cells \#1023
and \#1024 show distinct individual magnetic field signatures, consistent with
the division of the cell's surface area in two during mitosis. The resulting
change in magnetic field strength from $\approx \SI{3.9}{\mu\tesla}$ peak
amplitude of the mother cell is in line with the reduction of SPION density on
the daughter cell surface due to the mitosis, as during this process the mother
cell grows and separates into two equal daughter cells. This results in a
theoretical increase in surface area of \SI{26}{\percent}, causing a subsequent
reduction of SPION density of $\approx \SI{21}{\percent}$, which can be observed
in the magnetic field image with the resulting magnetic field peak amplitudes of
the daughter cells of \SI{3.1}{\mu\tesla} (\#1023) and \SI{2.2}{\mu\tesla}
(\#1024) respectively. We attribute the discrepancy to the deformed dipole
signature of cell \#1024 due to movement during the acquisition time. Cell
\#1023 shows a well-structured, symmetric dipole that better reflects the
absolute labeling strength. This demonstrates clearly how the information of the
occurring processes is encoded in the magnetic field information. Furthermore,
evidence of either tumor cell cannibalism \cite{Durgan2017} or entosis
\cite{Druzhkova2024,Krishna2016} is observed, as cells \#46 and \#135 merge into
cell \#411 between the \SI{40}{\hour} and \SI{48}{\hour} time points
(Fig.~\ref{fig:figure_3}b).  We attribute the faint magnetic field signature of
the resulting merged cell to a larger size, as well as a change in the distance
of the cell's center of mass and thus a change in the mean distance of SPIONs on
the surface of both involved cells away from the sensor surface. Additionally,
due to the coupling between the SPION position and the cell surface,
morphological changes of the cell are also observable in the magnetic field
channel (Fig.~\ref{fig:figure_3}b, c). Cell \#333 shows filopodia outgrowth,
characteristic for cancer cells, and Cell \#186 shows vertical stretching of its
cell membrane, which are both visible in the magnetic field image as a change in
the shape of the magnetic dipole signature, as the cell's shape is encoded in
the magnetic field information. This washing out of the magnetic dipole field is
attributed to the elongation of the cell as well as movement during the
\SI{5}{\min} integration window of the ODMR measurement. This subsequently
causes partial compensation of the magnetic dipole field, as the positive and
negative lobes of the dipole field cancel out over a larger area. Additionally,
the movement of the cell during this process washes out the magnetic field
signature, as the magnetic field affecting the NV centers is averaged over the
integration time of the measurement. These demonstrations illustrate the
capability of the magnetic field imaging modality to resolve single-cell
dynamics based on the change of magnetic field signatures induced by
magnetically labeled cells. As the resulting magnetic field of a cell is
composed of the cumulative contribution of individual SPIONs situated on the
cell surface (Eq.~\ref{eq:B}), changes in the cell's position and shape are
reflected in the magnetic field signatures. QCM thus provides a unique method to
observe fundamental cellular processes irrespective of optical accessibility or
necessary excitation doses incident on the sample, which would otherwise degrade
or interfere with the sample viability.

\subsection*{Magnetic imaging of cell population dynamics}
\begin{figure}[H]
    \centering
    \includegraphics{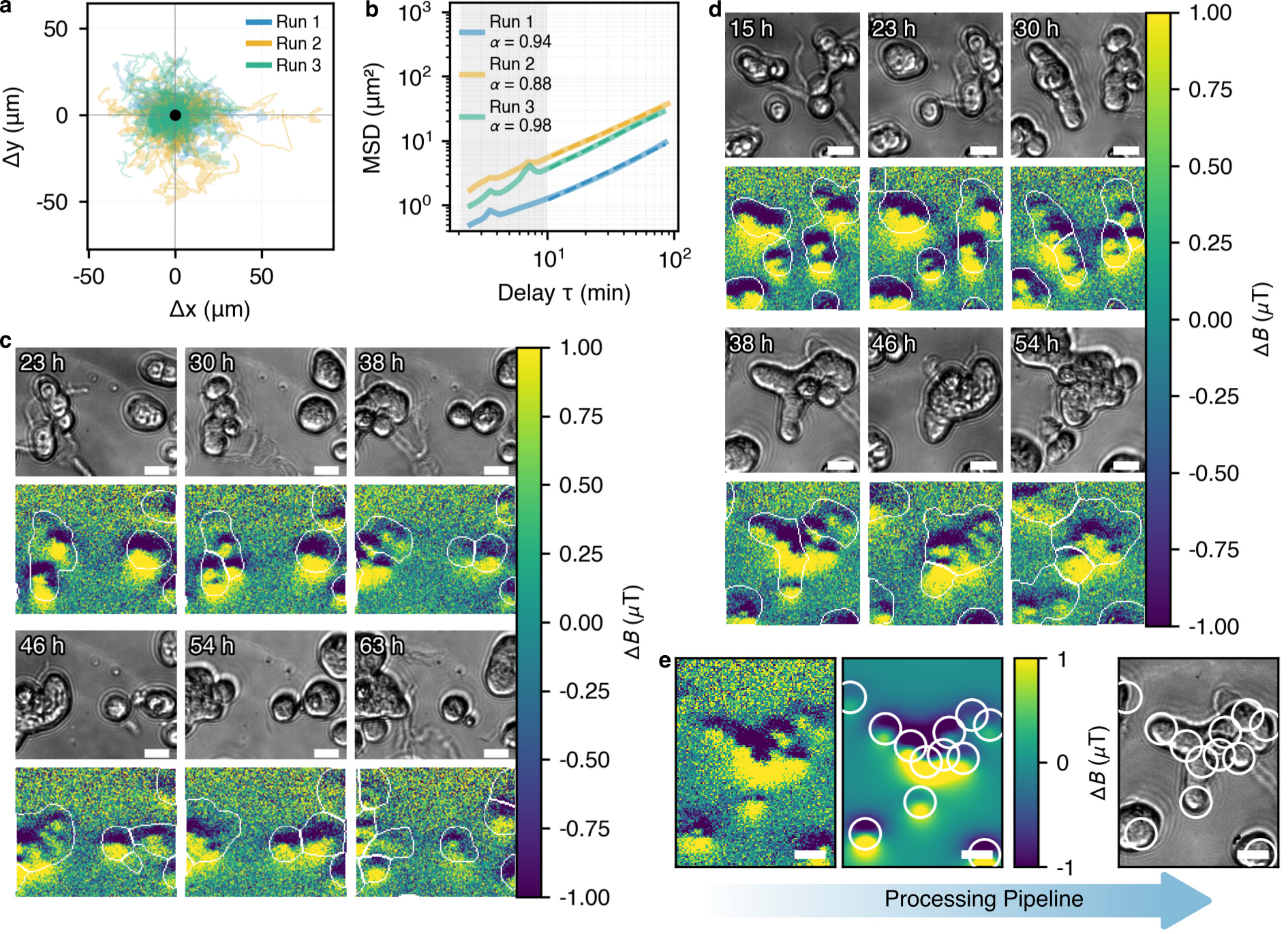}
    \caption{\textbf{Selected ROIs showing observable cell culture dynamics
    captured in correlated brightfield and magnetic field imaging of a
    \SI{66}{\hour} timespan experiment.} \textbf{a}, Spider plot showing the
    trajectories of cells that could be tracked for $>$\SI{6}{\hour}
    continuously. The trajectories are colored by experiment and illustrate that
    migration is limited to a few tens of \si{\micro\meter} over the course of
    the experiment. \textbf{b} The mean-squared displacement (MSD) of the
    tracked cells shows a slight sub-diffusive behavior ($\alpha < 1$), as
    expected for adherent single cells in a monolayer culture
    \cite{Hohmann2022}. Low brightfield image quality, segmentation, and
    tracking errors in the dense cell clusters cause significant centroid noise,
    which dominates for low $\tau$ values (shaded region). The MSD curve is thus
    evaluated for $\tau >$ \SI{10}{\minute}, where the MSD curve shows
    sub-diffusive behavior. \textbf{c}, A group of cells undergoing cluster
    formation, subsequent splitting and migration out of the FoV. Merging and
    separation of the cell cluster is clearly visible in magnetic field images.
    Cell IDs are omitted for visual clarity, as well as faulty segmentations due
    to the incoherent cluster structure posing a challenge for the segmentation
    algorithm. \textbf{d}, Complex cell interactions imaged with correlated
    imaging modalities. After formation of larger cell clusters individual cells
    remain partly visible in the magnetic field map, in contrast to the
    brightfield image. All scale bars: \SI{20}{\micro\meter}. \textbf{e}, Cell
    localization predicted from the magnetic field map, without any input from
    brightfield data. From left to right, the experimentally captured magnetic
    field map, the modeled field obtained by sparse dipole deconvolution and
    peak localization (cells of identical \SI{20}{\micro\meter} size, white
    circles), and the brightfield image overlaid with the magnetically recovered
    cell positions (white circles). Because the magnetic stray fields traverse
    biological matter without attenuation, individual cell identities are
    recovered from their dipole signatures alone, even within dense clusters
    where brightfield contrast collapses.}
    \label{fig:figure_4}
\end{figure}

In addition to single-cell behavior, a larger perspective is resolvable in the
magnetic field imaging in the form of cell-cell interactions and cluster
formation. Over the course of three experiments with identical conditions, but
different time scales (\SI{48}{\hour} for Run 1, \SI{66}{\hour} for Run 2, and
\SI{72}{\hour} for Run 3, not including the initial seeding time), the HT29
cells proliferate, migrate and form dense monolayer clusters as they move within
the FoV. To evaluate the migration of cells over the course of the experiment,
we analyzed the trajectories of individual cells whose tracks exceeded
\SI{6}{\hour} of continuous observation (Supplementary Sec.~S3) in the spider
plot (Fig.~\ref{fig:figure_4}a). It is evident that cells show limited migration
across the FoV, with most cells showing net-zero migration and only a few
outliers within 2-3 cell diameters ($\approx$\SI{20}{\micro\meter}). To further
evaluate the migratory behavior, we evaluated the mean-squared displacement
(MSD) $\langle r^2(\tau) \rangle = K_\alpha \tau^\alpha$ of the tracked cells
(Fig.~\ref{fig:figure_4}b), where $K_\alpha$ (Supplementary Sec.~S3) is the
diffusion coefficient, $\alpha$ is the anomalous diffusion exponent and $\tau$
the evaluated time delay between time points. Evaluation of the MSD was limited
to a delay $\tau \in \lbrack \SI{10}{\minute}, \SI{90}{\minute} \rbrack$ due to
centroid noise dominating for low $\tau$ caused by the segmentation, tracking
errors, and loss of cell tracking due to cluster formation. The upper limit of
$\tau$ is in line with the commonly applied criterion for limiting $\tau\leq
T_\mathrm{traj}$ where $T_\mathrm{traj} = \SI{6}{\hour}$ is the minimum
trajectory length \cite{Saxton1997,Michalet2010}. The resulting MSD curve shows
a slight sub-diffusive behavior, with $\alpha$ values between 0.88 and 0.98
across the experiments. This quantified sub-diffusive exponent ($\alpha<1$)
serves as a fundamental physical justification for our discrete sampling
approach. Because the structural constraints of the monolayer restrict the
maximum cellular displacement over \SI{8}{\hour} to a fraction of the average
intercellular distance, cell identity can be unambiguously tracked and
reconstructed purely from consecutive magnetic dipole maps. This directly
demonstrates that the magnetic channel is fully self-sufficient, rendering the
high-frequency optical co-registrations an overdetermined baseline used strictly
for independent ground truth verification. 

These processes are clearly visible in the magnetic field channel, allowing us
to track the dynamics of individual cells as they form, and interact with,
clusters of cells. Magnetic field imaging successfully resolves the grouping and
splitting of cell clusters, as well as their migration across the FoV during the
\SI{66}{\hour} experiment duration, based on their magnetic dipole signatures.
Fig.~\ref{fig:figure_4}c illustrates this as a group of cells (right side)
splits at the \SI{38}{\hour} time point, and one of the resulting subgroups
subsequently merges into another at the \SI{63}{\hour} time point. Additionally,
the magnetic field channel allows us to track individual cells within clusters,
while the brightfield image shows immediate loss of contrast, as well as merging
of cell boundaries at the \SI{38}{\hour} and \SI{46}{\hour} time points
(Fig.~\ref{fig:figure_4}d). Because the cells are all labeled with SPIONs, we
still resolve some degree of individuality in the magnetic field channel, while
the brightfield image shows a merged cluster of indistinguishable cells.
Fig.~\ref{fig:figure_4}c (left) and d (center) show the formation of a dense
cell cluster, whose details remain unresolvable in the brightfield modality, but
the respective magnetic field maps of the captured time points resolve details
of individual cells within the cluster. 

Building on this, the position and iron mass of individual cells can be
recovered directly from the magnetic field map alone, without any optical input
(Fig.~\ref{fig:figure_4}e). We demonstrate this by applying a lightweight
localization algorithm to the magnetic field map of Run~2 at the \SI{38}{\hour}
time point, where cells assembled to larger clusters. As a leading order
approximation that keeps the reconstruction numerically tractable, each cell is
modeled as a single point dipole of magnetic moment $\mathbf{m}$ magnetized
along the interrogated NV axis, which defines a kernel that is invariant under
translation (Eq.~\ref{eq:B}). We recover the underlying source distribution
through a sparse, nonnegative, regularized deconvolution of the field map with
the dipole kernel. This collapses each cell's extended dipole signature into a
single localized peak and suppresses sidelobe artifacts, so that the cell count
emerges directly from the deconvolution. A fixed minimum separation between
cells extracts the candidate centers, and a joint nonnegative least squares fit
of the superposed dipole fields eliminates residual false detections while
reading out the peak field amplitude of each cell. For this first order
approximation, holding the cell size fixed at its nominal value, this amplitude
is converted into a per cell iron mass through the known specific magnetization
of the SPIONs \cite{Pointner2025}, so that the procedure returns an estimation
of both position and iron mass across the ROI. We overlay the recovered
positions onto the brightfield image as circles of the chosen fixed size of
\SI{20}{\micro\meter} (Fig.~\ref{fig:figure_4}e), and direct comparison with the
brightfield image shows that the magnetic localization recovers the positions of
individual cells within the cluster. This proof of concept demonstrates that the
magnetic field map alone contains sufficient information to recover individual
cell identities, even within dense clusters where brightfield contrast
collapses. This is a key advantage of magnetic-field imaging over brightfield:
cell identity is preserved through cluster formation, a capability usually
associated with fluorescence labels.

This performance highlight uncovers the core advantage of QCM over
state-of-the-art optical modalities: when cellular boundaries completely
collapse in the brightfield channel due to local tissue density rendering
classical segmentation algorithms blind, the magnetic channel successfully
maintains individual cell identities. This sustained multi-day observation is
directly enabled by our specialized custom incubation system: while its
optimized optics isolate the excitation plane to minimize background noise, its
integrated micro-environmental control ensures a continuous, optimal supply of
nutrients and metabolic stability directly on the diamond substrate.
Consequently, QCM transcends the physical limits of photon scattering without
compromising long-term cell kinetics, proving its capability to operate as an
independent imaging modality where optical microscopy fundamentally fails.
Unlike fluorescence, however, magnetic-field imaging is unaffected by
photobleaching or scattering, enabling sustained monitoring over multi-day
timescales regardless of microenvironmental conditions of the sample
\cite{Yoon2020,Mela2024}.

\section*{Discussion}\label{sec:discussion} 

We demonstrate that single-cell information from living cells can be encoded in
stray magnetic fields and read out at subcellular resolution over multi-day
timescales. SPION based magnetic labeling combined with widefield NV
magnetometry inside a specialized incubator allows QCM to resolve
proliferation, mitosis, entosis, morphological remodeling, and cluster dynamics
in HT29 colorectal cancer cells over \SI{72}{\hour}, including inside dense
clusters of cells where brightfield contrast collapses. This capability
addresses a long-standing gap left by existing live cell imaging modalities.
Optical methods are bound by their information carrier and trade photon budget
against duration and depth resolution, meaning that multi-day fluorescence
imaging inevitably induces phototoxic cell stress and suffers from optical
scattering \cite{Yoon2020,Mela2024}. QCM fundamentally circumvents these
constraints by replacing optical photons with static magnetic stray fields as
the primary carrier of cellular information.

While widefield NV center magnetometry has previously been introduced for
single-cell evaluation, prior state-of-the-art approaches remained strictly
limited to static, single time endpoint measurements of fixed samples
\cite{LeSage2013,Mathes2024,Davis2018,Chen2022,Glenn2015}. This historical
limitation stems from an unresolved biophysical dilemma in requiring continuous,
high-power microwave and laser delivery, which inherently introduces severe
phototoxic and thermal stress, rendering multi-day live observation impossible
\cite{DelRosario2025,Icha2017}. Our platform overcomes this technological
bottleneck by introducing a self-contained architecture. By combining a
specialized incubation system directly with a carefully designed microscope
architecture, we decouple sustained spin readout from environmental and
phototoxic degradation of the sample. This architectural synergy transforms
diamond magnetometry into a fully functional, longitudinal live cell imaging
modality capable of operating under true physiological conditions.

Consequently, the physical parameters mapped by QCM provide a robust framework
for monitoring long-term cell kinetics, specifically cell migration, without the
need for optical accessibility. Adherent single cells within dense biological
assemblies typically exhibit complex, sub-diffusive migratory trajectories that
are severely obscured or entirely lost in brightfield imaging once clusters form
\cite{Plou2018}. Our tracking platform successfully addresses this challenge by
preserving absolute cell identity through continuous cluster formation, fusion,
and mitotic splitting based purely on shifting magnetic dipole signatures.
Crucially, these multi-hour reconstructions serve as a quantitative validation
of this spatial framework. 

Because static magnetic stray fields traverse biological matter without loss at
cellular length scales, QCM bypasses both optical scattering and attenuation
within magnetically translucent environments, such as soft tissue
\cite{Shapiro2004,Cheng2023,Heyn2006}. At current SPION labeling strengths,
individual cell signatures remain resolvable several tens of \si{\micro\meter}
above the sensor surface based on extrapolation and an SNR of 1 (Supplementary
Sec.~S4). Advancements in the magnetic moment and size tuning of SPION labels
offer the potential to at least double the saturation magnetization of these
nanoparticles \cite{Heinen2026}. Additional tuning of SPIONs to form clusters
further amplifies the per-cell signal strength, due to more iron per EpCAM
binding site, which directly scales the depth at which single cells remain
trackable via QCM, extending the depth range to and beyond
\SI{100}{\micro\meter} (Supplementary Sec.~S4) \cite{Pointner2025}. 

It is important to emphasize that the \SI{8}{\hour} sampling interval for
magnetic field image capture used in this study does not represent a fundamental
technical lower limit of the QCM hardware. Rather, it was strategically chosen
as an operational optimum tailored to the slow baseline kinetics of HT29 cells
while strictly minimizing cumulative laser exposure, which makes QCM a
well-suited choice for imaging slow cell dynamics. For future applications in
highly dynamic environments or dense 3D tissue models where optical
co-registration is entirely unavailable, the magnetic sampling frequency can be
scaled up significantly. This scalability is governed by the underlying sensor
physics, where the optical NV-readout is shot-noise limited and the
signal-to-noise ratio scales as SNR $\propto B_{c} \cdot \sqrt{\Delta t_{int}}$,
with $B_{c}$ being the local cell magnetic field strength and $\Delta t_{int}$
the ODMR integration time. Implementing high-moment SPION clusters addresses
this in two distinct regimes. For dynamic cellular events occurring within the
first dozens of micrometers away from the sensor, the amplified magnetic field
$B_{c}$ drastically reduces the required integration window. Since the
integration time scales quadratically with the field strength for a constant
target SNR ($\Delta t_{int} \propto 1/B_{c}^2$), a conservative 10-to-20-fold
increase in local magnetization reduces the necessary integration time from the
current \SI{5}{\minute} down to less than a second. Due to this exponential drop
in laser exposure per frame, the actual measurement interval between individual
scans can be compressed from \SI{8}{\hour} down to minutes while maintaining a
lower cumulative phototoxic dose, enabling high-frequency, optics-free
monitoring. Second, for cells situated deeper within dense 3D architectures
($\approx\SI{100}{\mu \meter}$) where migration is naturally slowed down by
microenvironmental constraints, the enhanced magnetic moment ensures that
individual signatures remain resolvable above the sensor's noise floor (SNR
$\geq 1$), sustaining robust longitudinal monitoring over multi-day timescales
even under the sparse sampling regime. 

Beyond extending the imaging range and improving the imaging rate, the
reconstruction itself offers room for refinement. Conceivable improvements
include a more extensive biologically-informed forward model that represents
cells as extended sources to capture their 3D shape and expresses the magnetic
field through a spherical harmonic (multipole) expansion rather than point
dipoles, more advanced sparse deconvolution schemes for this underdetermined
inverse problem, and machine learning approaches that directly learn the mapping
from magnetic field maps to cell position, shape, and iron mass information
\cite{Chen2022,Pointner2025}.

Building on this capability, the presented monitoring approach establishes a
versatile technological baseline, demonstrating that single-cell kinetic
pathways can be continuously resolved over dozens of hours under physiological
conditions forming the foundation for adaptation into complex 3D tissue
architectures, which is especially relevant with rising interest in 3D culture
models such as organoids and tumor spheroids
\cite{Jassin2025,Edmondson2014,Dubois2017} (Supplementary Sec.~S5). This
technological baseline directly opens up avenues for advanced oncology research
that previously had to rely on retrospective inferences from fixed histology or
employ imaging modalities that ultimately trade phototoxic dose, label
photostability and penetration depth against observation time. Where previous
experimental research relies on multiphoton or confocal imaging to investigate
the process of cancer cell invasion, QCM would resolve the individual cells
invading and migrating deep inside dense stroma tissue over multi-day
timescales, capturing the invasion dynamics that current methods only infer from
snapshots \cite{Provenzano2008,Gaggioli2007}. Especially for imaging of cancer
cell dynamics in 3D environments, where confocal microscopy is insufficient to
resolve the full depth of the sample, QCM would allow continuous monitoring of
the sample in contrast to light sheet microscopy, which requires imaging of
independent samples for each time point due to terminal mounting procedures
\cite{Bozkurt2021}. As single cell models do not capture the full complexity of
the tumor microenvironment, experiments using spheroids or organoids are
increasingly used to study cancer cell invasion. Recent studies investigate the
development of such micro-tumors using light sheet microscopy, an imaging
modality that necessitates imaging of independent systems for subsequent time
points \cite{Diosdi2025}. Furthermore, current research in cell invasion based
on spheroids reduces the observation to a 2D equatorial plane of the organoid
relying on local-maximum detection instead of cell segmentation, which is
insufficient to capture the full 3D dynamics of the system \cite{Prikryl2026}.
QCM would allow continuous monitoring of the same organoid over multiple days,
resolving individual labeled cells deep within the sample, further improving the
quality of fundamental biological research.

Ultimately, these milestones bridge a long-standing gap at the intersection of
quantum physics and the life sciences. For biomedical imaging, QCM introduces a
paradigm where subcellular spatial resolution and multi-day continuous
observation no longer come at the cost of phototoxic cell stress, remaining
fully functional even within optically dense cellular environments. For quantum
sensing, this work demonstrates that non-invasive magnetic stray fields can be
successfully translated into quantitative biophysical models of geometry and
monitoring kinetics. QCM firmly establishes magnetic field-encoded imaging as a
powerful, self-contained modality for the future of translational cancer
research.

\section*{Methods}\label{sec:methods}
\subsection*{Widefield NV-center magnetic field microscope}
The widefield magnetic field microscope is built around a $2 \times 2 \times
0.5$~\si{\cubic\milli\meter} [100] bulk diamond with a \SI{1}{\micro\meter}
thick, \SI{1}{ppm} NV-doped surface layer with [$^{15}$N] isotopic enrichment
\cite{quantum-diamonds}. The diamond is mounted on a polished glass cube using
optical adhesive (NOA61, Norland) to enable TIR excitation of the NV layer,
preventing the \SI{532}{\nano\meter} laser (Verdi G Series, Coherent) from
interacting with the biological sample \cite{Pointner2025incubator}. The laser
was focused to illuminate the FoV without clipping on the sample edges. The
power was then tuned to \SI{25}{\milli\watt} over the \SI{376}{\micro\meter}
diameter FoV, which was found to be a good compromise between magnetic field
sensitivity and phototoxicity over the \SI{5}{\minute} integration duration of a
single ODMR measurement. Two samarium cobalt ring magnets provide a static bias
field of up to \SI{40}{\milli\tesla} along the selected NV axis to magnetize
SPIONs present in the sample \cite{Pointner2025}. NV fluorescence is collected
by a long working distance objective (MY50X-805, Mitutoyo), filtered by a
\SI{635}{\nano\meter} long-pass filter (EdgeBasic) and imaged onto a $4\times4$
binned electron multiplying charge-coupled device camera sensor with a
$1024\times1024$ pixel sensor (Evolve13, Teledyne). A dual-tone microwave
signal, generated by IQ mixing a dual-tone intermediate frequency from an
arbitrary wave-form generator (OPX+, Quantum Machines) with a swept local
oscillator (SynthNV Pro, Windfreak), drives both $^{15}$N hyperfine transitions
of the NV ground state \cite{Schloss2018}. The center frequency of the microwave
signal is swept over \SI{4}{\mega\hertz} intervals across the
$\ket{0}\rightarrow\ket{-1}$ and $\ket{0}\rightarrow\ket{+1}$ transitions. The
camera was subsequently exposed for \SI{200}{\milli\second} with and without the
microwave signal to compute the spin contrast. The resulting ODMR contrast is
evaluated per pixel to extract the electron Zeeman splitting and subsequently
the local magnetic field per pixel, yielding spatially resolved magnetic field
map over a \SI{266}{\micro\meter}~$\times$~\SI{266}{\micro\meter} field of view.
Background subtraction using a rolling-ball algorithm removes low-frequency
spatial noise, isolating the dipole signatures of individual SPION-labeled cells
\cite{Glenn2015}. 

\subsection*{Incubation system for live-cell imaging}
To maintain cell viability during extended imaging sessions, a specialized
incubation platform compatible with the NV widefield microscope geometry was
employed \cite{Pointner2025incubator}. The incubation chamber consists of an
acrylic cylinder bonded to the glass cube with optical adhesive (NOA61,
Norland), sealed to prevent leakage of culture medium. Temperature control is
provided by regulated aluminum heating elements (MP825-20.0-1\%, Caddock) with
negative temperature coefficient thermistor feedback, calibrated to maintain an
incubation temperature of \SI{37}{\degreeCelsius} at the diamond surface. The
incubator lid is heated \SIrange{2}{3}{\degreeCelsius} above the incubation
temperature to suppress condensation. A premixed gas atmosphere
(\SI{5}{\percent} CO$_2$, \SI{95}{\percent} air; 2GF Mixer, Okolabs) is sensed
via a CO$_2$ sensor and subsequently humidified (CO2-500ml, Bioscience Tools)
before entering the 3D printed chamber (BioMed White Resin, Formlabs). An
internal water reservoir and overflow system maintain stable humidity to prevent
medium evaporation and pH drift. CO$_2$ exchange with the culture medium is
facilitated through grooves on the incubation volume lid, while a sealed
cultivation compartment minimizes evaporative loss. All hardware components are
positioned outside the incubation atmosphere to avoid contamination and
interference with the ODMR measurement. Further details of the incubator design
are provided in \cite{Pointner2025incubator}. Brightfield imaging of the cells
is performed by a light emitting diode illuminating the sample from above and
collecting the transmitted light through the imaging objective on the camera.

\subsection*{Cell segmentation and tracking}
Brightfield time-series images were segmented using Cellpose
\cite{Stringer2021}, a deep-learning-based generalist algorithm for cellular
segmentation. The cell diameter was fixed to \SI{70}{px}, flow threshold and
cell probability threshold were set to 0.6 and -2 respectively based on
empirical testing. The resulting instance masks were linked across frames using
btrack \cite{Ulicna2021}, which employs a Bayesian multi-object tracking
algorithm to reconstruct cell trajectories and detect proliferation events. A
$\lambda$ branching of 50, prospected segmentation miss rate of
\SI{10}{\percent}, search radius of \SI{50}{px}, optimization tolerance of
\SI{0.5}{\percent} was used for the tracking. Together, the pipeline provides
per-cell position, lineage, and morphological information from the brightfield
channel, which is spatially correlated with the corresponding magnetic field
maps. Due to the limited contrast and resolution of the brightfield images
acquired through the NV microscope objective, segmentation accuracy is
insufficient for quantitative morphological analysis. Cell outlines are
therefore displayed as visual guides to facilitate spatial correlation between
the brightfield and magnetic field channels.

\subsection*{Magnetic field cell localization}
To recover cell positions from a magnetic field map after background
subtraction, without optical input, each cell is approximated as a point
magnetic dipole oriented along the interrogated NV $\lbrack 111 \rbrack$ axis.
The NV projected stray field of a unit dipole (Eq.~\ref{eq:B}) defines a single
kernel $K$ that is invariant under translation, so the map is modeled as
$B_{c,\parallel} \approx K * x$ with $x \geq 0$ a sparse map of dipole
amplitudes per pixel. We estimate $x$ by minimizing $\tfrac{1}{2}\lVert K * x -
B_{c,\parallel}\rVert_2^2 + \lambda \lVert x \rVert_1$ subject to $x \geq 0$
with a fast iterative shrinkage thresholding algorithm (FISTA) \cite{Beck2009}.
The $L_1$ penalty enforces sparsity, collapsing each cell's bipolar signature
into a single peak and suppressing the sidelobe artifacts that a plain matched
filter ($K^{\mathsf{T}} B$) produces. Cell centers are extracted from $x$ by
nonmaximum suppression with a fixed minimum separation and an edge margin, and a
single joint nonnegative least squares fit of the superposed dipole fields
removes residual spurious candidates and yields the peak field amplitude of each
cell. Holding the cell size fixed at its nominal value, this amplitude is mapped
to a per cell iron mass through the known specific magnetization of the SPIONs,
calibrated by the peak field of a single dipole that carries a reference iron
mass at the nominal cell depth \cite{Pointner2025}. An optional subpixel
refinement of the $(x,y)$ position of the brightest sources further sharpens the
localization. The detection noise floor is estimated from the median absolute
deviation of the discrete Laplacian of the map. On synthetic ground truth data
the recovered positions reach an accuracy of $\approx\SI{0.2}{\micro\meter}$
root mean square.

\subsection*{Cell culture and SPION labeling}
Prior to cell seeding, the polished diamond surface was functionalized with PLL
to promote cell adhesion \cite{Glenn2015}. The incubation volume above the
diamond was filled with \SI{100}{\micro\liter} of PLL solution and incubated
for \SI{1}{\hour} at \SI{37}{\degreeCelsius}. The PLL solution was subsequently
aspirated, and the surface was allowed to dry under sterile conditions before
cell seeding. 

The mycoplasma-free human colorectal cancer cell line HT29 was
cultivated at \SI{37}{\degreeCelsius} and \SI{5}{\percent} CO$_2$ in RPMI-1640
medium (Gibco Life Technologies, \#P04-18500) supplemented with
\SI{10}{\percent} fetal bovine serum (PAN Biotech, \#P30-3031), penicillin
(100~U/mL) and streptomycin (100~U/mL) (PAN Biotech, \#P06-07100). Cells were
used at \SIrange{70}{80}{\percent} confluence, between passages 3 and 13.

For labeling, cells were detached from the plate by trypsinization. $1 \times
10^{6}$ cells were resuspended in \SI{1}{\milli\liter} of cultivation medium.
All subsequent incubation steps were performed at \SI{37}{\degreeCelsius} on a
shaker at 300~rpm. Between incubation steps, the cell solution was centrifuged
($300 \times g$ for \SI{5}{\minute}) to remove excess antibody or nanoparticles
and washed with Dulbecco's phosphate buffered saline (DPBS, PAN Biotech,
\#P04-36500). Cells were first incubated with anti-EpCAM antibody (1:250
dilution in cultivation medium; Cell Signaling Technology, \#2929) for
\SI{45}{\minute}, followed by a biotinylated secondary goat-anti-mouse antibody
IgG H\&L (1:500 dilution in cultivation medium; Abcam, \#ab6788) and
streptavidin-coated superparamagnetic iron oxide nanoparticles (1:2 dilution in
DPBS; Ocean NanoTech, \#SHS-20) for \SI{30}{\minute} each.

\subsection*{Monolayer imaging protocol}
To image cell migration and proliferation in a monolayer culture, HT29 cells
were seeded at a concentration of 40\,000 cells per \SI{500}{\mu\liter} culture
medium directly onto the PLL-coated diamond surface. After allowing $\approx$
\SI{8}{\hour} for cell adhesion, the first ODMR measurement was acquired.
Subsequent ODMR measurements were performed at \SI{8}{\hour} intervals, while
brightfield images were acquired every \SIrange{2}{5}{\minute} to track cell
migration and proliferation dynamics. The culture medium was refreshed with
prewarmed medium after each ODMR acquisition to maintain nutrient supply and
optimal culture conditions.

\section*{Data availability}\label{sec:data-availability} The presented datasets
generated during the current study are available from the corresponding authors
upon reasonable request.

\section*{Code availability}\label{sec:code-availability} The custom code used
for data acquisition and analysis in this study is available from the
corresponding authors upon reasonable request.

\backmatter

\bmhead{Supplementary information}

\bmhead{Acknowledgements}
We thank Lukas Heinen and the team of the Section of Experimental Oncology and
Nanomedicine (SEON) at the University Hospital Erlangen for fruitful discussions
and support on the physical and biochemical aspects of superparamagnetic
iron-oxide nanoparticles. We are grateful to Daniela Thalheim and Jessica
Knittel of the Chair of Experimental Tumor Pathology for their support with cell
and tissue culture, labeling, and sample preparation. We further thank Prof. Dr.
Jörg Wrachtrup, Prof. Dr. Ronald Walsworth, Dr. Durga B. R. Dasari, Dr. Amit
Finkler and Dr. Yaron Antebi for fruitful discussions and editorial input.

\bmhead{Funding}
A.P. and R.N. disclose support for the research of this work from the European
Union under the Key Digital Technologies Joint Undertaking (KDT JU), now Chips
Joint Undertaking (Chips JU), in the projects ARCHIMEDES (grant nos. 16MEE0329
and 101112295) and MOSAIC (grant nos. 16MEE0494 and 101081238), and from the
German Federal Ministry of Education and Research (BMBF, now BMFTR) in the
projects INNOBAT (grant no. 03XP0492D) and QuaLiProM (grant no. 03XP0573C).
Regine Schneider-Stock was supported by the Deutsche Forschungsgemeinschaft
(DFG) SCHN477/19-1 and BMBF (QUBIS).

\bmhead{Author contributions}
A.P. designed and built the experimental setup, performed measurements, data
analysis and prepared the figures. A.P. and R.N. conceived the study. S.B. and
D.T. prepared cell cultures and SPION labeling protocols. R.S.-S. supervised the
biological experiments and contributed to data interpretation. A.P. and R.N.
wrote the manuscript with input from all authors.

\bmhead{Competing interests}
A.P., R.S.-S., R.N. are inventors on a patent application related to the
optical excitation and collection scheme described in this work (application
number EP 24198554.8 and EP 10 2025 109 562.0) 

\bibliography{references}

\end{document}


\begin{center}
  {\Large \textbf{Supplementary Information}}\\[0.5em]
  {\large Multi-day single-cell magnetic imaging of living cancer cells using Quantum Cell Monitoring}\\[1em]
  Andre Pointner, Sarah Belasi, Regine Schneider-Stock, Roland Nagy
\end{center}

\setcounter{section}{0}
\renewcommand{\thesection}{S\arabic{section}}
\renewcommand{\thefigure}{S\arabic{figure}}
\renewcommand{\thetable}{S\arabic{table}}
\renewcommand{\theequation}{S\arabic{equation}}

\section{NV magnetometry compatible incubation platform}
\label{sec:incubator}

\begin{figure}[H]
  \centering
  \includegraphics{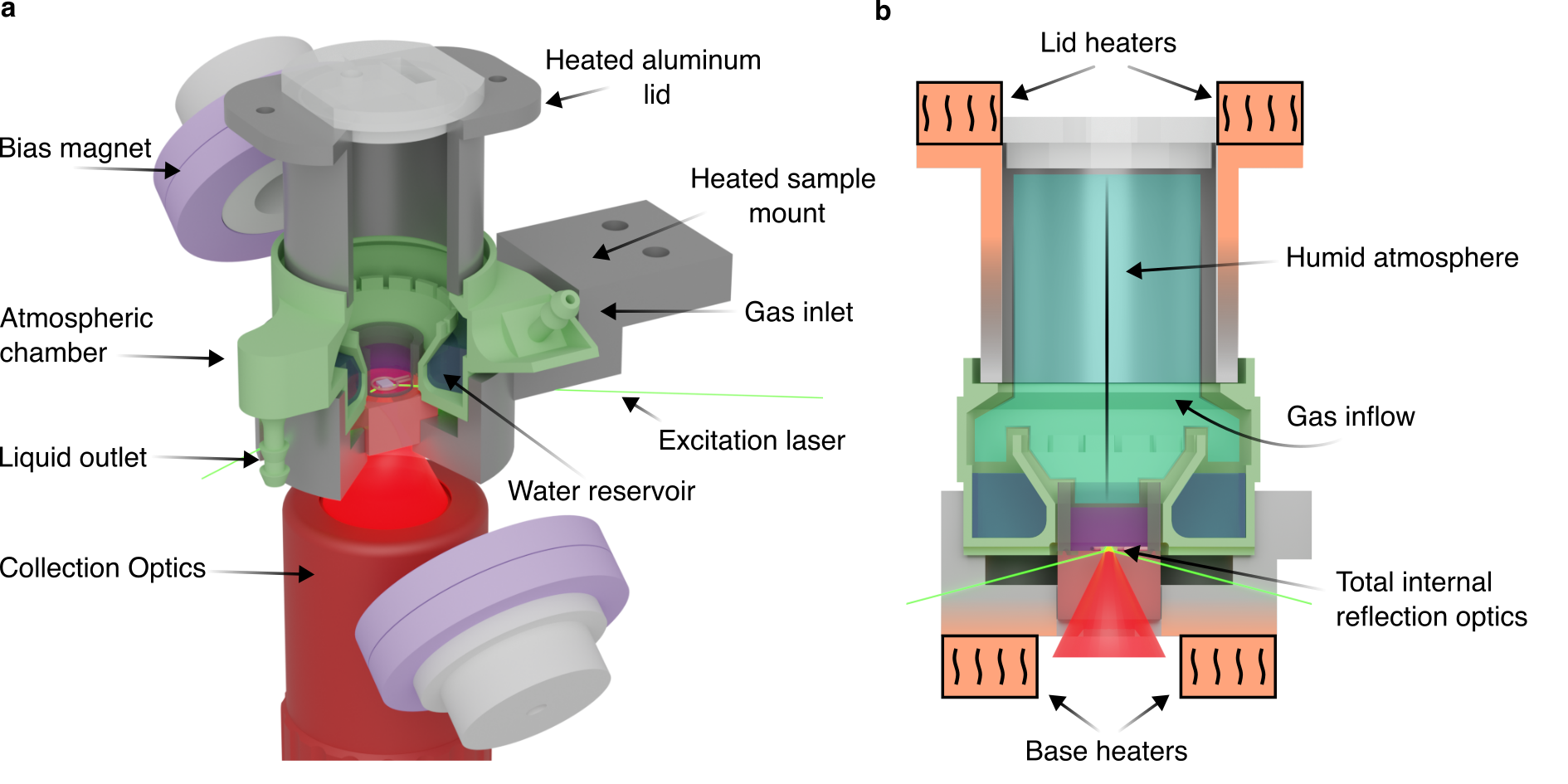}
  
  \caption{\textbf{Integrated incubation system on the NV widefield microscope.}
  \textbf{a}, Render of the incubation assembly, showing the atmospheric chamber
  (green) and lid (gray) enclosing the incubation volume between the two
  samarium-cobalt bias magnets (purple). The heated sample mount stages the
  glass cube required for total internal reflection (TIR), on which the diamond
  sample is mounted. The acrylic culture cylinder is bonded to the polished
  glass cube, same as the diamond sample and microwave antenna. The excitation
  laser is incident from below and reflects symmetrically out of the assembly.
  The optical access from below allows collection of NV emission onto the camera
  optics. The gas inlet, condensate overflow and water reservoir are integrated
  in the 3D printed atmospheric chamber. \textbf{b}, Cross-section of the
  incubation chamber enclosing the cultivation volume within the acrylic
  cylinder (gray). The water reservoir provides thermal and humid buffering,
  while the gas inflow supplies a constant stream of water and the premixed
  incubation gas, resulting in a stable incubation volume (turquoise shading).
  Two pairs of resistive heaters heat the lid and the sample stage base to
  maintain the incubation temperature at \SI{37}{\celsius}. The excitation laser
  is cleanly reflected at the diamond--medium interface, while the NV emission
  is collected through the optical path from below.}
  \label{fig:incubator_render}
\end{figure}

Long-term cultivation directly on the diamond sensor surface requires an
incubation environment compatible with the inverted total internal reflection
(TIR) illumination geometry, microwave access and permanent magnets capable of
reaching $\geq \SI{25}{\milli\tesla}$, which precludes the use of available
commercial stage-top or whole-microscope incubators \cite{Pointner2025incubator}
(Fig.~\ref{fig:incubator_render}a). The incubation chamber is constructed from
an acrylic cylinder bonded to the polished glass cube via optical adhesive
(NOA61, Norland) \cite{LeSage2013}, defining a small cultivation volume of
\SI{500}{\micro\liter} directly above the diamond surface. The cylinder is
enclosed by a 3D-printed atmospheric chamber, manufactured from a biocompatible
resin (BioMed White Resin, Formlabs) on an SLA printer (Form 4, Formlabs). The
chamber is printed as a single body in which the gas-inlet and
condensate-overflow hose connectors are integrated directly into the resin part,
minimizing the amount of metal introduced near the sensing region and the
magnetizing bias field and yielding a drop-in replaceable component that
requires no post-processing. Its internal volume is partitioned into an outer
reservoir, which holds the water that sustains the humidified atmosphere and
acts as a thermal buffer, and an inner incubation volume that seals over the
open-top of the acrylic cylinder (Fig.~\ref{fig:incubator_render}b). The full
assembly fits between the samarium-cobalt bias magnets and around the microwave
antenna without compromising ODMR sensitivity, and all active hardware, namely
the heaters, thermistors, and gas and humidity supply, is kept outside the
incubation atmosphere to avoid contamination of the culture and interference
with the ODMR measurement \cite{Pointner2025incubator}.

A premixed \SI{5}{\percent}~\ce{CO2} / \SI{95}{\percent} air atmosphere is
supplied to the chamber, with the \ce{CO2} fraction set by metering the \ce{CO2}
mass flow into a gas mixer (2GF Mixer, Okolabs) against the ambient air
delivered by an air pump. The gas mixture is humidified in a
temperature-controlled humidifier (CO2-500ml, Bioscience Tools) before entering
the chamber, where the internal water reservoir and a condensate overflow
maintain saturated humidity to suppress evaporation of the culture medium. A
diffusion lid with peripheral grooves separates the cultivation volume from the
surrounding humidified atmosphere, permitting \ce{CO2} exchange while
suppressing medium evaporation and the associated pH drift. The \ce{CO2}
concentration is monitored by a non-dispersive infrared (NDIR) sensor
(ExplorIR-20W) placed upstream of the humidifier, since elevated humidity
degrades the sensor, and is logged at \SI{2}{\hertz} over a USB-TTL serial
interface alongside all thermistor readings. Slow drift of the \ce{CO2}
concentration, arising from variations in ambient pressure, temperature, and
supply-gas pressure, is corrected by periodically readjusting the \ce{CO2} flow
at intervals no longer than the \SI{8}{\hour} spacing between ODMR acquisitions,
at which point the culture medium is also replenished
\cite{Pointner2025incubator}.

Temperature is regulated by two heated aluminum elements driven in independent
parallel PID loops by a microcontroller (Arduino Nano) using thick-film
resistors (MP825-20.0-1\%, Caddock) and NTC thermistors as temperature sensors.
The base heater is the primary element: it mounts the glued glass-cube, diamond,
and microwave-antenna stack together with the incubation chamber via a nylon set
screw, provides optical access from below for fluorescence collection, and
leaves lateral gaps for the TIR excitation, heating the incubation volume from
the substrate side for a faster and more direct thermal response than heating
the lid alone. The heated lid carries an identical pair of resistive heaters and
thermistor and is closed by a transparent cover that permits brightfield
illumination and visual monitoring of the culture-medium level. It is operated
\SI{2}{\celsius} to \SI{3}{\celsius} above the incubation temperature to
suppress condensation on the optical surfaces. To avoid placing a sensor inside
the cultivation volume, where it would perturb the ODMR measurement, the
gradient between the external base-heater sensor and the diamond surface was
calibrated against an in-medium reference sensor, yielding a stable offset of
\SI{3}{\celsius}. The heater set point of \SI{34}{\celsius} therefore maintains
the diamond surface at \SI{37}{\celsius} while bounding the worst-case
temperature excursion below physiological limits in the event of thermistor
beta-value drift. The temperature is monitored at four points, namely the base
heater, the lid, the incubation atmosphere, and the ambient laboratory
environment, and a control measurement confirmed that the applied microwave
power produces no measurable heating of either deionized water or culture
medium, so the microwave drive does not perturb the incubation temperature
\cite{Pointner2025incubator}. Surface functionalization of the diamond with
Poly-L-Lysine (PLL) ensures rapid cell adhesion prior to the start of multi-day
ODMR acquisition.

\begin{figure}[H]
  \centering
  \includegraphics{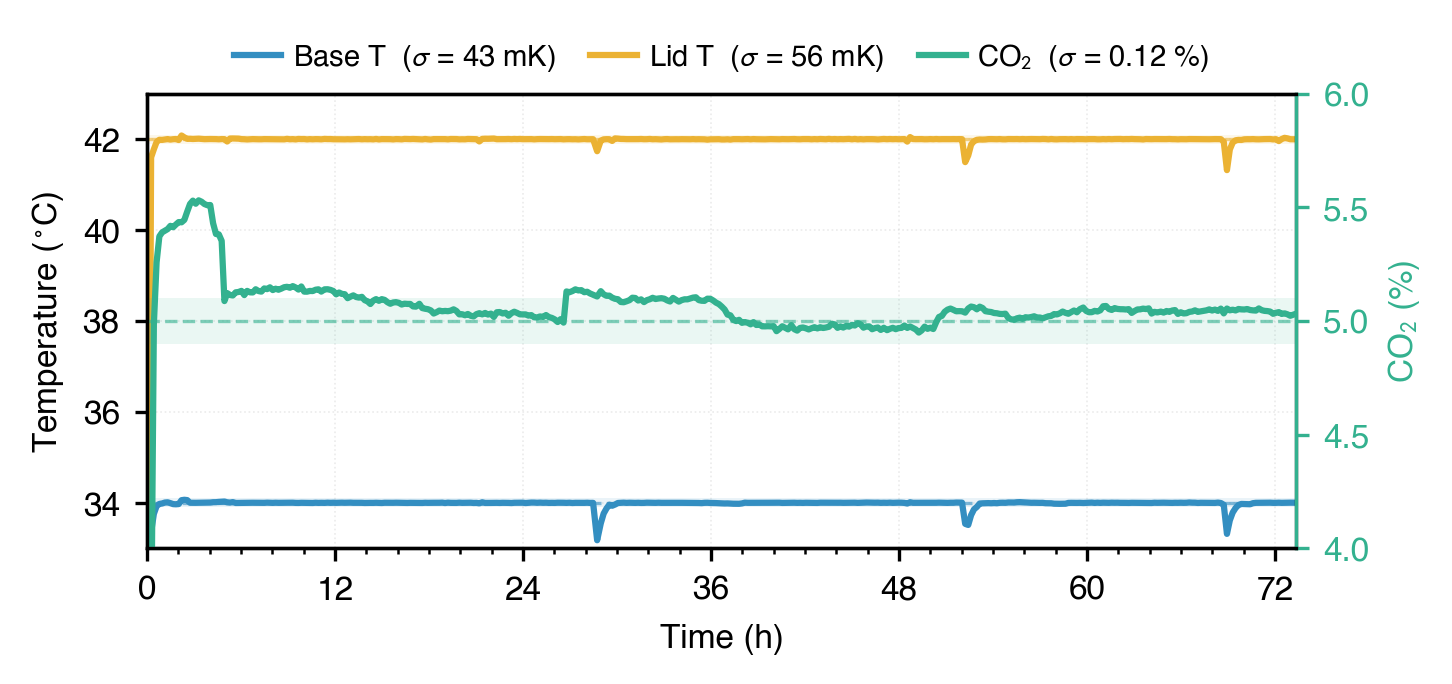}
  \caption{\textbf{Long-term temperature and \ce{CO2} stability of the
  incubation system.} Base- and lid-heater temperatures and \ce{CO2}
  concentration recorded over a \SI{72}{\hour} incubation period of Run 3 from
  the main manuscript. The temperature traces show a steady state for the
  majority of the experiment ($\sigma_T = \SI{43}{\milli\kelvin}$ base,
  \SI{56}{\milli\kelvin} lid) and the $\approx\SI{0.12}{\percent}$ \ce{CO2}
  drift. The temperature traces show semi-periodic drops in temperature, which
  are attributed to the periodic refreshment of the culture medium, requiring
  opening the incubator lid.}
  \label{fig:incubator_stability}
\end{figure}

Starting from room temperature, the Arduino-based PID-controlled heaters reach
their set points within approximately \SI{30}{\minute} and subsequently hold the
surface at the target temperature, with residual fluctuations of $\sigma_T =
\SI{43}{\milli\kelvin}$ at the base and $\SI{56}{\milli\kelvin}$ at the lid,
i.e.\ below \SI{100}{\milli\kelvin} and within the control range of the loop.
Over a \SI{72}{\hour} period the \ce{CO2} concentration varies by only
$\approx\SI{0.12}{\percent}$, confirming that the system maintains a stable
physiological environment across multi-day experiments
(Fig.~\ref{fig:incubator_stability}). To verify that this environmental control,
rather than the medium or surface functionalization, is responsible for the
observed cell viability, a stress test was performed in which HT29 cells were
cultured for \SI{48}{\hour} with the temperature and \ce{CO2} regulation
disabled \cite{Pointner2025incubator}. Under these conditions the cells showed
no proliferation, adhesion, or migration and exhibited membrane blebbing
characteristic of early apoptosis, demonstrating that the controlled incubation
environment is essential for sustaining the cultures imaged in this work.

\section{Cell damage due to high laser flux}
\label{sec:cell_damage}

\begin{figure}[H]
  \centering
  \includegraphics{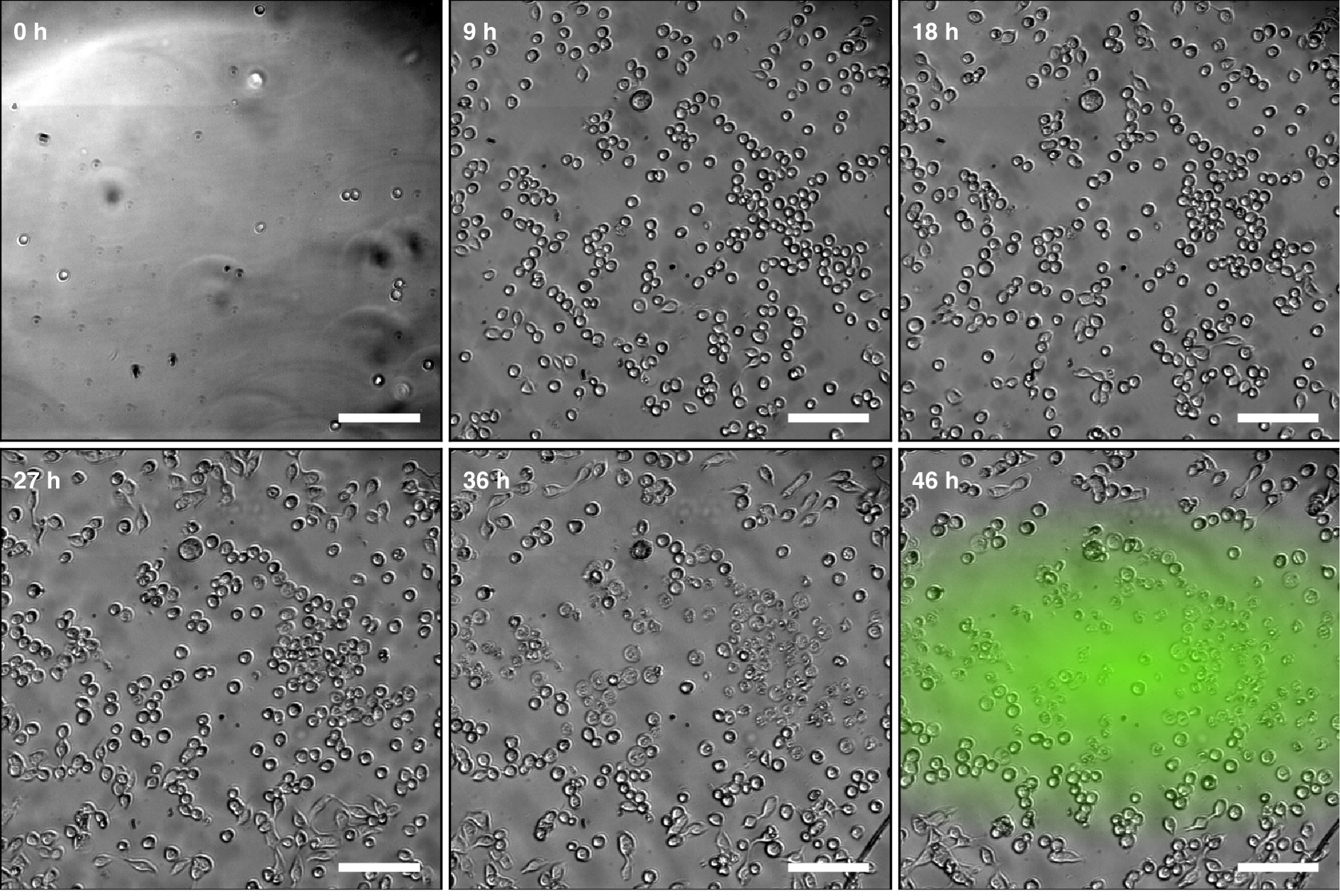}
  \caption{\textbf{Cell damage due to high laser flux.} Timeseries of a
  \SI{48}{\hour} experiment with periodic ODMR measurements every
  \SI{30}{\minute}. The full FoV of $\approx$\SI{650}{\micro\meter} is imaged
  with a $20\times$ magnification. The green shaded area illustrates the area of
  the Gaussian excitation beam that is totally reflected by the diamond surface.
  Significant cell damage due to the excitation laser is observable within the
  area of the beam spot, while cells at the edges of the field of view show
  healthy morphology and proliferation behavior.}
  \label{fig:cell_damage}
\end{figure}

Repeated ODMR acquisition over multi-day experiments imposes a cumulative
optical dose on the biological sample that, if not carefully constrained, leads
to severe phototoxic damage. To characterize the operating regime within which
cell viability is preserved, we first monitored a HT29 monolayer over
\SI{48}{\hour} under high-flux ODMR acquisition every \SI{30}{\minute}, imaging
the full $\approx$\SI{650}{\micro\meter} FoV at $20\times$ magnification
(Fig.~\ref{fig:cell_damage}).

During the initial \SI{8}{\hour} settling period, cells adhered uniformly to the
PLL-functionalized diamond surface across the full FoV. Beyond $t
\approx$\SI{27}{\hour}, cells within the totally reflected Gaussian beam spot
(green overlay) exhibited progressive loss of morphological integrity and
accumulation of cellular debris. In contrast, cells at the periphery, where the
local laser flux is significantly lower, retained a healthy state and continued
to proliferate and migrate. By $t \approx$\SI{36}{\hour}, no viable cells
remained within the central beam region. The spatial coincidence between the
damaged region and the beam footprint immediately implies excessive optical flux
as the dominant damage source.

Three mechanisms could in principle account for this damage: laser-induced
heating of the diamond surface, microwave-induced heating of the culture medium,
or direct phototoxicity from the excitation laser. Laser-induced heating of the
diamond is excluded by the absence of any measurable temperature rise at the
surface under comparable conditions \cite{Pointner2025incubator}. The diamond's
high thermal conductivity would in any case dissipate heat quickly across the
sample, rather than produce the beam-localized damage observed here.
Microwave-induced heating of the medium is excluded by the antenna geometry, the
wire surrounds the field of view, so medium heating would be strongest at the
FoV periphery, yet cells closer to the wire remain adherent and viable while
damage occurs in the FoV center, far from the antenna. Only phototoxicity, which
acts locally where the laser intensity is highest, matches the observed spatial
distribution \cite{Matsumoto2014,Waeldchen2015}.

\begin{figure}[H]
  \centering
  \includegraphics{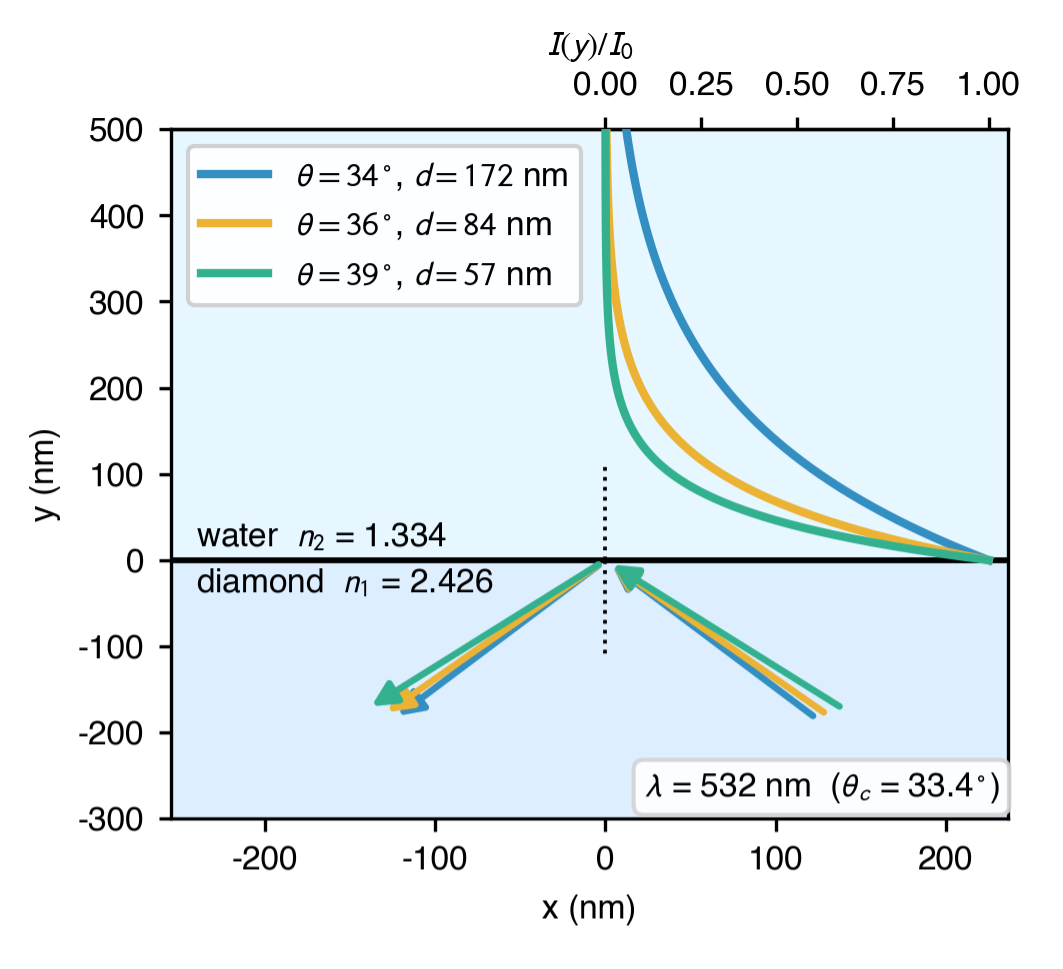}
  \caption{\textbf{Simulated evanescent-field amplitude at the diamond--water
  interface.} Simulation of the evanescent wave coupling into water for a
  \SI{532}{\nano\meter} excitation wavelength from diamond (n=2.426) into water
  (n=1.334) for different incidence angles $\theta$. The evanescent wave
  penetrating from the total internal reflection barrier into the culture medium
  or cell surface is governed by an exponential decay with distance from the
  interface \cite{Axelrod1981}. However, this still corresponds to a significant
  exposure to the excitation laser for the cells in close proximity to the
  diamond surface.}
  \label{fig:evanescent_wave_sim}
\end{figure}

Although the total internal reflection (TIR) geometry confines the
\SI{532}{\nano\meter} excitation beam within the diamond and is intended to
protect cells from direct exposure \cite{LeSage2013}, a residual evanescent
field penetrates into the culture medium with an exponential decay characterized
by a sub-wavelength penetration depth \cite{Axelrod1981}. Simulation of
evanescent-wave coupling into water shows that, within the first
\SI{100}{\nano\meter} of the surface, cells are exposed to a significant
fraction of the excitation laser intensity (Fig.~\ref{fig:evanescent_wave_sim}).
This near-surface exposure, compounded by scattering from residual surface
imperfections, accounts for the localized damage observed along the beam path.
Reflection of the laser away from the sample volume, even for a perfectly flat
interface, is therefore insufficient to prevent phototoxic effects. The effect
of stray scattered light from the sample holder causes additional optical dose
incident on the cells.

\begin{figure}[H]
    \centering
    \includegraphics{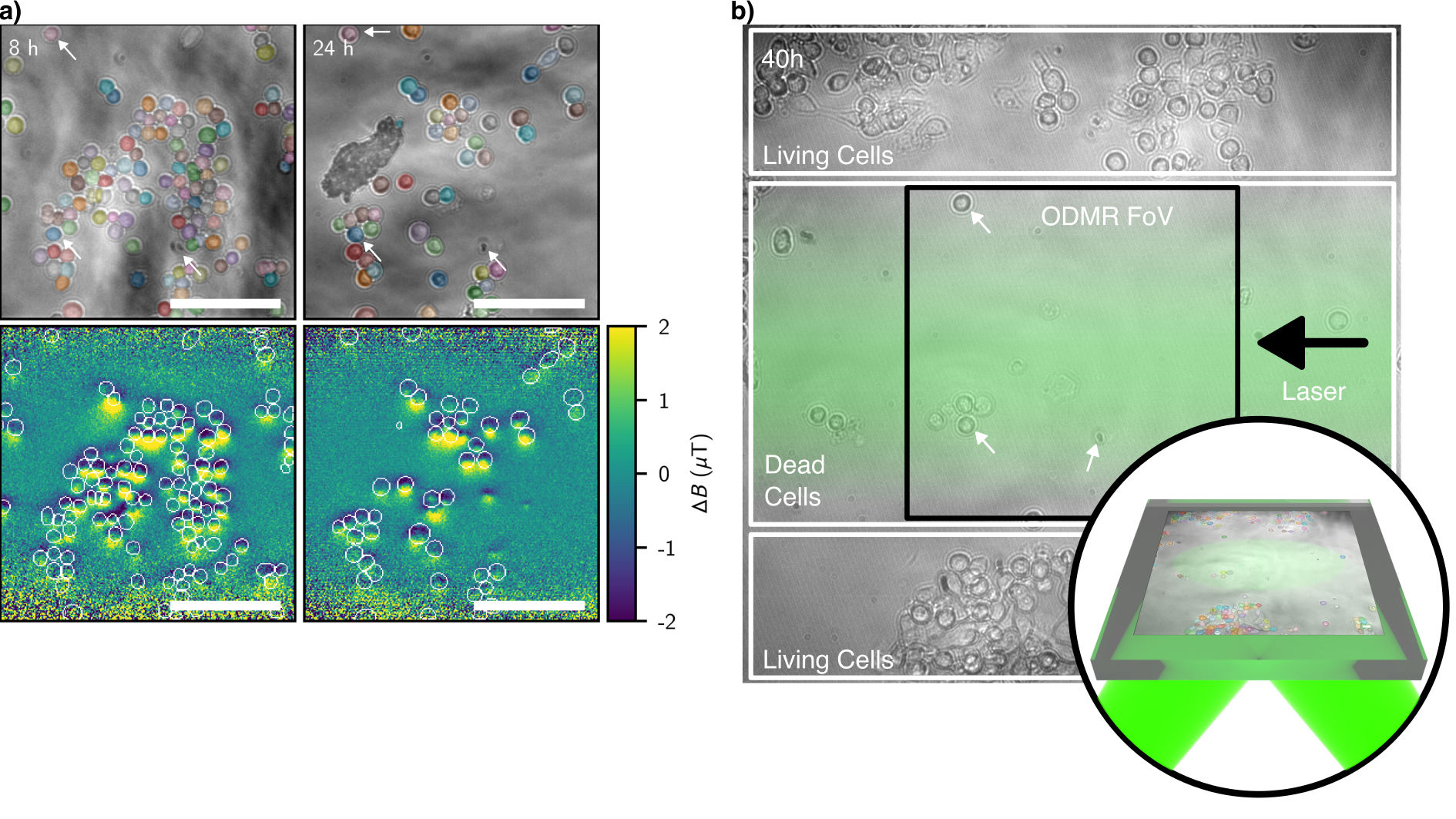}
    \caption{\textbf{Phototoxicity-induced cell detachment during multi-day ODMR
    acquisition.} \textbf{a}, Correlated brightfield (top) and magnetic field
    (bottom) images of SPION-labeled HT29 cells at \SI{8}{\hour} (left) and
    \SI{24}{\hour} (right) showing characteristic magnetic dipole signatures of
    individual cells. Colored overlays in brightfield and white outlines in
    magnetic field images indicate cell boundaries segmented with Cellpose
    \cite{Stringer2021}. White arrows mark cells that persist across both time
    points. \textbf{b}, Full FoV brightfield image at \SI{40}{\hour}. Black
    square encloses the ODMR evaluation region; cell tracking failed due to the
    detachment of the majority of cells from photodamage; black arrow and green
    shaded area illustrate the direction of the excitation laser beam path;
    white boxes enclose regions of cell adhesion lateral to the beam and cell
    detachment along the beam path. Inset: three-dimensional rendering of the
    total internal reflection (TIR) excitation geometry, illustrating the laser
    beam entering from below and reflecting at the diamond--medium interface.
    The light green overlay indicates the Gaussian beam spot at the TIR plane.}
    \label{fig:figure_2}
\end{figure}

To mitigate photodamage, we increased the interval between ODMR measurements to
\SI{8}{\hour} to allow for cellular recovery. Even under these conditions,
cumulative photodamage is not fully eliminated over multi-day acquisition. Over
a \SI{40}{\hour} observation period, the majority of cells along the excitation
beam path progressively detach from the diamond surface, while cells lateral to
the beam remain adherent and proliferate (Fig.~\ref{fig:figure_2}). This
residual cell loss persists despite an \SI{8}{\hour} recovery interval between
ODMR measurements, demonstrating that the recovery aids in cellular survival,
but is not enough to ensure long-term viability under the current acquisition
parameters.

The operating point was not selected arbitrarily or through a purely empirical
sweep, but fixed by a sensitivity requirement set by the expected signal.
Saturated SPION labeling produces a per-cell stray-field signature of
\SI{1}{\micro\tesla} to \SI{2}{\micro\tesla}, consistent with both the labeling
density and the dipole amplitudes observed in our magnetic field maps. The
acquisition time and laser power must therefore yield a per-pixel sensitivity
whose noise floor lies well below this signal, so that individual cells are
resolved at a practical signal-to-noise ratio (SNR). For our \SI{5}{\minute}
acquisition, in which the dual-tone microwave is stepped across 102 frequency
points each imaged with and without microwave driving (204 frames of
\SI{100}{\milli\second} exposure) and repeated over 14 successive sweeps, the
total signal-integration time is $T_\mathrm{meas} \approx \SI{286}{\second}$.
With a per-pixel sensitivity of $\eta \approx
\SI{2.5}{\micro\tesla\per\sqrt\hertz}$, this gives a theoretical noise floor of
$\sigma_B = \eta/\sqrt{T_\mathrm{meas}} \approx
\SI{2.5}{\micro\tesla\per\sqrt\hertz}/\sqrt{\SI{286}{\second}} \approx
\SI{0.15}{\micro\tesla}$. This noise floor lies roughly an order of magnitude
below the expected cell signal, corresponding to an SNR of $\approx 7$ to $14$,
and thus leaves sufficient leeway to reliably resolve the magnetic field
signatures of individual cells. This balance defines the operating point:
lowering the laser power would further benefit cell viability but raise the
noise floor and degrade imaging quality, whereas raising it would improve the
SNR only marginally, since the cells are already detected reliably, at the cost
of increased phototoxic load. We therefore operate at \SI{25}{\milli\watt} over
the \SI{5}{\minute} acquisition within our $\varnothing$\SI{376}{\micro\meter}
FoV, the minimum optical dose that still guarantees robust single-cell
detection. The laser intensity, measurement duration, and recovery interval are
thus not free parameters but jointly constrained by the interplay of magnetic
field sensitivity and cell viability.

\section{Cell migration tracking}
\begin{figure}[H]
  \centering
  \includegraphics{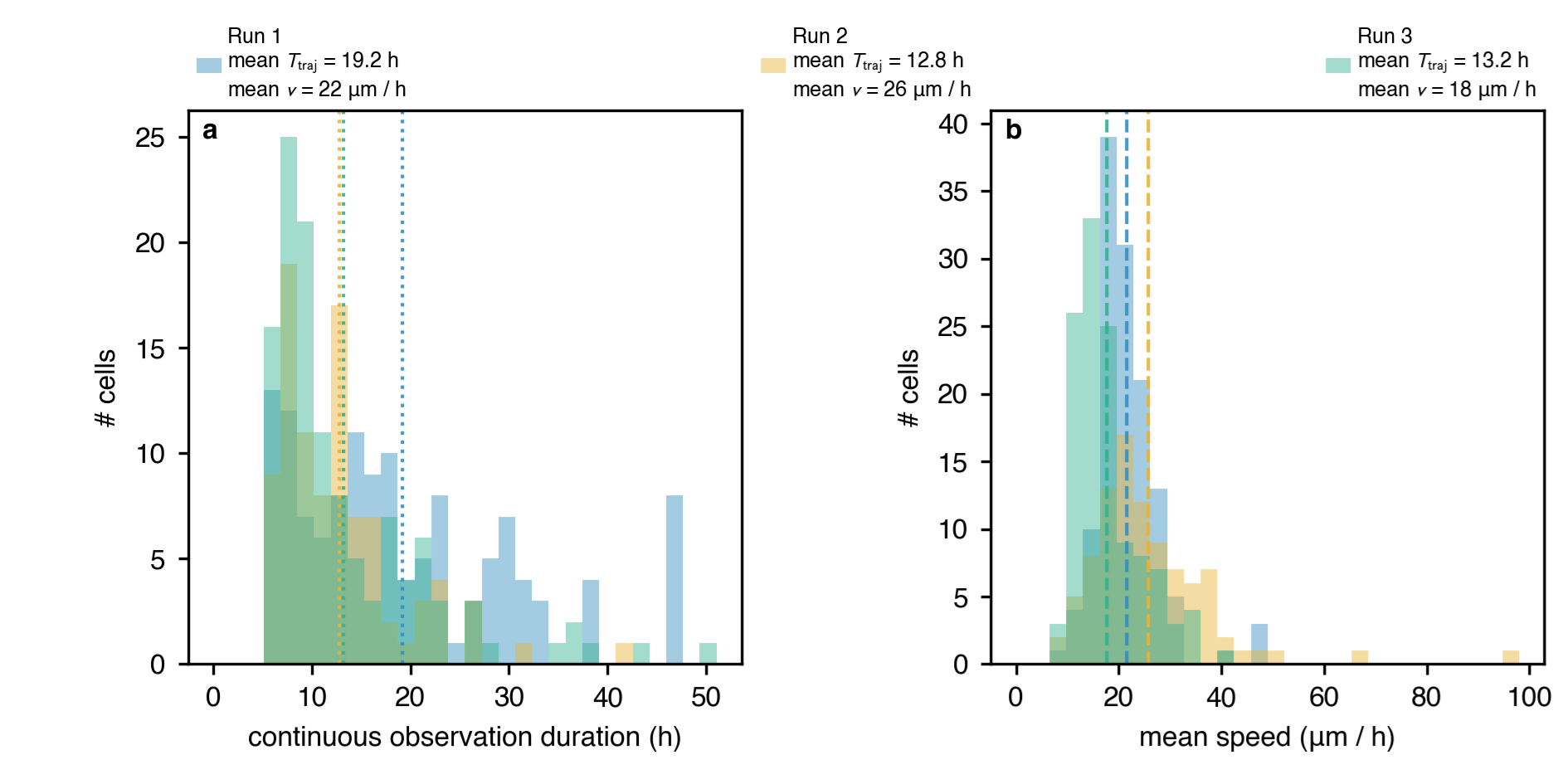}
  \caption{\textbf{Observation duration and speed distribution of tracked HT29
  cells.} \textbf{a}, Histogram of observation durations for cells in the
  long-term tracking dataset. Evaluated cell trajectories are clipped above
  \SI{6}{\hour} to filter out short, non-persistent tracks in order to provide a
  clearer picture about cell migration during the experiment. The average
  trajectory lengths of the three experiment runs are \SI{19.2}{\hour} (Run 1),
  \SI{12.8}{\hour} (Run 2), and \SI{13.2}{\hour} (Run 3) respectively. The steep
  decline in the number of tracks with increasing duration can be attributed to
  tracking errors and segmentation faults, that make segmenting and tracking of
  individual adjacent cells difficult. \textbf{b}, Histogram of average cell
  speeds for the tracked cells in the long-term tracking dataset. The average
  cell speeds across the three experiment runs are
  \SI{22}{\micro\meter\per\hour} (Run 1), \SI{26}{\micro\meter\per\hour} (Run
  2), and \SI{18}{\micro\meter\per\hour} (Run 3) respectively. The distribution
  of cell speeds shows a mean around \SI{20}{\micro\meter\per\hour} for all
  three runs.}
  \label{fig:cell_track}
\end{figure}

Evaluating the migration behavior of the tracked cells is crucial for
understanding the dynamics of the cell population and the underlying biological
processes. By analyzing the trajectories of the tracked cells, we can extract
information about their migration patterns, such as their speed, directionality,
and persistence. The histogram of observation durations for the tracked cells in
the long-term tracking dataset shows a steep decline in the number of tracks
with increasing duration, which is why we clipped the trajectories above
\SI{6}{\hour} continuous observation (Fig.~\ref{fig:cell_track}a). This can be
attributed to cell clustering, segmentation errors, and tracking issues during
the experiment. The average trajectory length of the three experiment runs are
\SI{19.2}{\hour} (Run 1), \SI{12.8}{\hour} (Run 2), and \SI{13.2}{\hour} (Run 3)
respectively. The histogram of average cell speeds for the tracked cells in the
long-term tracking dataset shows a mean around \SI{20}{\micro\meter\per\hour}
for all three runs (Fig.~\ref{fig:cell_track}b). The mean-square displacement
(MSD) analysis of the cell trajectories shows a sub-diffusive behavior, which is
characteristic of cells migrating in a crowded environment. The resulting MSD
fits return $K_{\alpha,1} =
\SI{0.136}{\micro\meter\squared}\,\mathrm{min}^{-\alpha_1}$ and $\alpha_1 =
0.94$ for Run 1, $K_{\alpha,2} =
\SI{0.706}{\micro\meter\squared}\,\mathrm{min}^{-\alpha_2}$ and $\alpha_2 =
0.88$ for Run 2, and $K_{\alpha,3} =
\SI{0.372}{\micro\meter\squared}\,\mathrm{min}^{-\alpha_3}$ and $\alpha_3 =
0.98$ for Run 3 respectively. The distribution of the cell speed should be
interpreted with caution, as highly mobile cells are more likely to not be part
of a larger cluster and thus are more likely to be tracked, while less mobile
cells are more likely to be part of a larger cluster and thus are more difficult
to track.

\section{SNR-based depth limit of single-cell monitoring}
\label{sec:snr_limit}

\begin{figure}[H]
  \centering
  \includegraphics{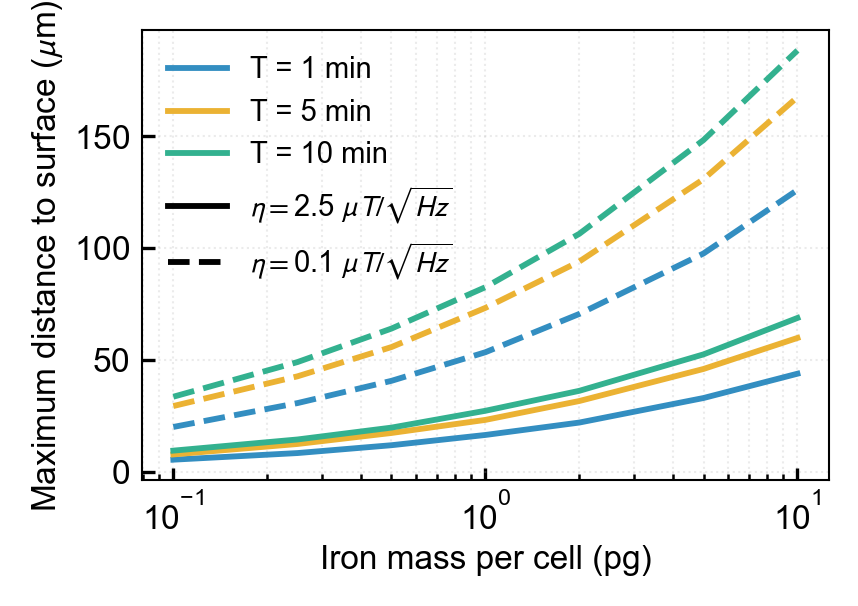}
  \caption{\textbf{Simulated detection horizon for individual SPION-labeled
  cells.} Maximum standoff $d_\mathrm{max}$ at which the peak axial stray field
  $B_{c,\parallel}$ of a single labeled HT29 cell crosses the per-pixel noise
  floor $\sigma_B = \eta / \sqrt{T_\mathrm{meas}}$, as a function of iron mass
  per cell. Colors denote the ODMR integration time $T_\mathrm{meas}$ and line
  styles denote the per-pixel sensitivity, $\eta =
  \SI{2.5}{\micro\tesla\per\sqrt\hertz}$ for the present live-cell widefield
  setup (solid) and $\eta = \SI{0.15}{\micro\tesla\per\sqrt\hertz}$ for an ideal
  shielded configuration (dashed). The cell is modeled as a
  \SI{20}{\micro\meter} sphere with uniformly distributed SPIONs on its surface;
  $B_{c,\parallel}$ is extracted as the maximum of the simulated axial field map
  across a sensing plane at standoff $z$ below the cell.}
  \label{fig:snr_limit}
\end{figure}

To determine the standoff over which individual SPION-labeled cells remain
resolvable in Quantum Cell Monitoring (QCM), we performed a forward simulation
of the cellular magnetic stray field and compared its peak amplitude with the
per-pixel noise floor of the ODMR readout, following the framework introduced in
our previous SPION-labeling study \cite{Pointner2025}. A cell is approximated as
a sphere of diameter $d_c = \SI{20}{\micro\meter}$ whose surface carries the
SPION load, in line with the homogeneous distribution of SPIONs on the cell
surface observed by fluorescence microscopy \cite{Pointner2025}. The
cell-surface labeling is approximated by $N = 500$ points uniformly distributed
on the sphere, with each point acting as a perfect magnetic dipole carrying the
average moment of the SPIONs assigned to it. For a given iron mass per cell
$m_\mathrm{Fe}$, the number of bound particles follows from the mass-to-moment
relation of the streptavidin-coated Ocean NanoTech SHS-20 SPIONs
(single-particle saturated moment $m_p = \SI{8.6e-19}{\ampere\meter\squared}$ at
\SI{40}{\milli\tesla} \cite{Pointner2025}), and the cluster moment scales
linearly with the number of particles, because the bias field used here drives
the particles deep into the Langevin saturation region (\cite{Pointner2025}).
All dipole moments are aligned along the external bias field, which lies along
the chosen NV axis at the magic angle of $\approx 54.7^\circ$ to the diamond
surface normal. The cumulative magnetic field $B_z(x,y)$ in the sensing plane at
standoff $z$ below the cell is computed as the sum of the individual dipole
projections onto this NV axis, $B_z = \sum_k \mathbf{B}_k \cdot
\hat{n}_\mathrm{NV}$, on a $32 \times 32$ grid covering the cell footprint
\cite{Pointner2025}.

The peak amplitude $B_{c,\parallel}$ of the cellular signature is then read off
the simulated field map as $B_{c,\parallel}(z, m_\mathrm{Fe}, d_c) = \max_{x,y}
\abs{B_z(x,y)}$. At the magic angle, the dipole moments project onto the NV axis
as the characteristic positive/negative double-lobe pattern. The maximum of this
pattern approximately follows the amplitude returned by a 2D Gaussian-derivative
fit, the analytic form commonly used to extract per-cell stray-field amplitudes
from experimental $B_z$-maps \cite{Glenn2015}. The peak-pick is therefore the
simulative equivalent of the experimental fit and avoids introducing a
parametric model on a noiseless map. A monotonic decay curve
$B_{c,\parallel}(z)$ is generated at a reference iron mass $m_\mathrm{ref}$ and
linearly rescaled by $m_\mathrm{Fe}/m_\mathrm{ref}$ to cover the full sweep,
since the cell stray field at fixed diameter and distance is proportional to the
total magnetic moment of the surface-bound SPIONs \cite{Pointner2025}. For each
combination of iron mass and integration time the maximum detectable distance
$d_\mathrm{max}$ is then defined by the crossing
$B_{c,\parallel}(d_\mathrm{max}) = \sigma_B = \eta / \sqrt{T_\mathrm{meas}}$,
located on the rescaled curve by linear interpolation between adjacent
standoffs.

Iron loading was swept across $m_\mathrm{Fe} = \SIrange{0.1}{10}{\pico\gram\per
cell}$, spanning the labeling regime reported in our SPION optimization study
(saturating around \SI{2.7}{\pico\gram} per cell at
\SI{500}{\micro\gram\per\milli\liter} SPION concentration \cite{Pointner2025}),
and integration times spanned $T_\mathrm{meas} = $ \SIrange{1}{10}{\minute},
covering both the \SI{5}{\minute} dwell used here and longer settings accessible
with reduced phototoxic load. Two per-pixel sensitivities were considered: $\eta
= \SI{2.5}{\micro\tesla\per\sqrt\hertz}$, representative of the present
widefield setup under live-imaging laser power, and $\eta =
\SI{0.15}{\micro\tesla\per\sqrt\hertz}$, the sensitivity attainable on the same
hardware when the phototoxic dose on the cells is resolved, e.g., by a
reflective coating that fully shields the culture from stray light
(Fig.~\ref{fig:snr_limit}). The resulting curves are monotonic in
$m_\mathrm{Fe}$ and in $T_\mathrm{meas}$, so that improvements in either SPION
moment or integration time translate directly into increased trackable depth
\cite{Heinen2026}.

\section{Spheroid labeling for cell population dynamics}
\label{sec:spheroid}

\begin{figure}[H]
  \centering
  \includegraphics{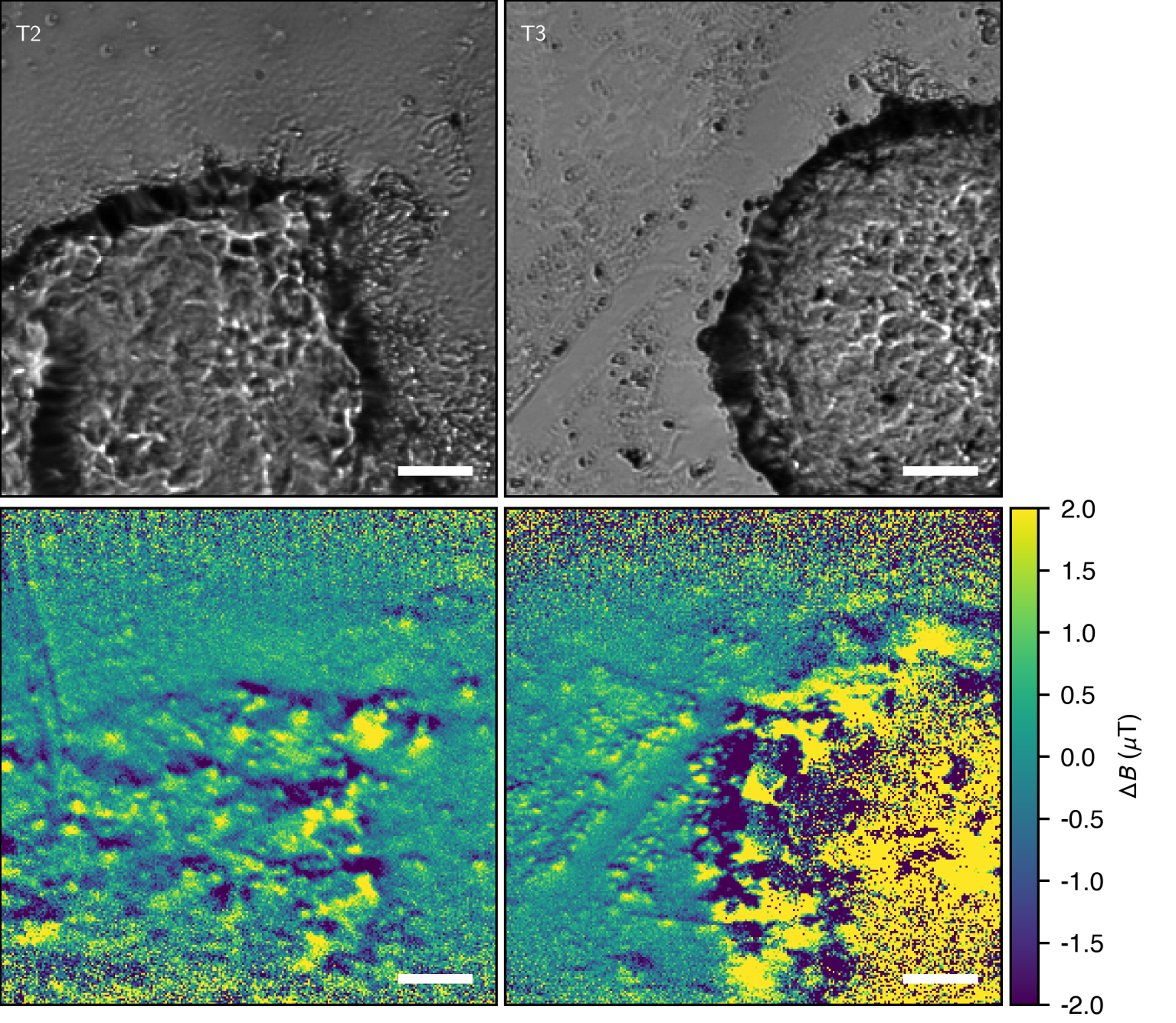}
  \caption{\textbf{Brightfield and magnetic field images of magnetically labeled
  HT29 spheroids.} \textbf{a}, Correlated images of a HT29 spheroid. The seed
  cell population of the spheroid was magnetically labeled with the same
  protocol as described in the main manuscript. The brightfield image shows
  discrete edges and clear structure of the cell spheroid. The magnetic field
  image reveals the underlying individual cells that remained magnetically
  labeled, even after extended proliferation periods. \textbf{b}, Correlated
  images of a HT29 spheroid with continuous re-labeling of the outer cells
  during growth. A significant jump in labeling strength of the outer cells is
  observable in the magnetic field image. The cumulative magnetic field strength
  of the spheroid cells closest to the diamond layer diminish the ODMR signal
  quality, due to a strong electron Zeeman shift outside of the sampled
  frequency range. Stray individual cells and cell residue is clearly visible
  outside of the spheroid, which can be attributed to dead cells or cell debris,
  that still carry some SPIONs.}
  \label{fig:spheroid}
\end{figure}

Beyond the single-cell level, QCM can be applied to track the dynamics of cell
populations. A promising approach is the labeling of spheroids, which are
three-dimensional aggregates of cells that represent tumor relevant
microenvironments. By magnetically labeling the seed cell population of the
spheroid or alternatively by continuously re-labeling the outer cells during
growth, a strong magnetic signal can be achieved due to the high cell density of
the spheroid. Imaging the spheroid under live-sustaining conditions allows for
monitoring its growth and migration over time. Within the scope of this work, we
have successfully labeled and imaged HT29 spheroids
(Fig.~\ref{fig:spheroid}), but have not yet performed a detailed analysis of the
spheroid dynamics. However, the preliminary results demonstrate that the
approach shows clear magnetic field signatures that open up the possibility to
track the growth and migration of the spheroid over time within living tissue.

\bibliographystyle{unsrt}
\bibliography{supp_references}